\documentclass{ar-1col-S2O}
\usepackage[comma]{natbib}
\usepackage{url}
\usepackage{graphicx}
\usepackage{epsfig}
\usepackage{natbib}
\usepackage{multirow}
\usepackage{amsmath}
\usepackage{amssymb}
\usepackage{subcaption}

\newcommand{\hi}{H\thinspace{\sc i}}
\newcommand{\hii}{H\thinspace{\sc ii}}
\newcommand{\hei}{He\thinspace{\sc i}}
\newcommand{\heii}{He\thinspace{\sc ii}}
\newcommand{\oi}{[O\thinspace{\sc i}]}
\newcommand{\oii}{[O\thinspace{\sc ii}]}
\newcommand{\oiii}{[O\thinspace{\sc iii}]}
\newcommand{\sii}{[S\thinspace{\sc ii}]}
\newcommand{\mgii}{Mg\thinspace{\sc ii}}
\newcommand{\cii}{C\thinspace{\sc ii}}
\newcommand{\ciii}{C\thinspace{\sc iii}]}
\newcommand{\civ}{C\thinspace{\sc iv}}
\newcommand{\sitwo}{Si\thinspace{\sc ii}}
\newcommand{\oip}{O\thinspace{\sc i}}
\newcommand{\lya}{Ly$\alpha$}
\newcommand{\fesc}{$f_{\rm esc}^{\rm LyC}$}
\newcommand{\fescrel}{$f_{\rm esc,rel}$}
\newcommand{\fesclya}{$f_{\rm esc}^{{\rm Ly}\alpha}$}
\newcommand{\sigsfr}{$\Sigma_{\rm SFR}$}
\newcommand{\cf}{$C_f$}
\newcommand{\logten}{$\log_{\rm 10}$}
\newcommand{\fratio}{$f_{\rm 900}/f_{\rm 1100}$}
\newcommand{\eg}{e.g.,}
\newcommand{\nhi}{$N_{\rm HI}$}
\newcommand{\Msol}{M$_\odot$}
\newcommand{\Zsol}{Z$_\odot$}
\newcommand{\araa}{Annu. Rev. Astron. Astrophys.}
\newcommand{\aj}{Astron. J.}
\newcommand{\aap}{Astron. Astrophys.}
\newcommand{\apj}{Ap. J.}
\newcommand{\apjl}{Ap. J. Lett.}
\newcommand{\apjs}{Ap. J. Suppl.}
\newcommand{\mnras}{MNRAS}
\newcommand{\nat}{Nature}

\jname{Xxxx. Xxx. Xxx. Xxx.}
\jvol{AA}
\jyear{YYYY}
\doi{10.1146/((please add article doi))}

\begin{document}

\markboth{Jaskot}{LyC Escape from Low-Redshift Galaxies}

\title{Ionizing Radiation Escape from Low-Redshift Galaxies and Its Connection to Cosmic Reionization}

\author{Anne E. Jaskot$^1$
\affil{$^1$Department of Physics and Astronomy, Williams College, Williamstown, MA, USA, 01267; email: anne.jaskot@williams.edu}}

\begin{abstract}
The escape of Lyman continuum (LyC) radiation from early galaxies transformed the intergalactic medium (IGM) and is intimately connected to the fueling and feedback processes that regulate galaxy evolution. IGM attenuation interferes with high-redshift LyC observations, but growing samples of LyC observations at $z<0.1$ are revealing the properties of LyC-emitting galaxies. Along with multi-wavelength observations of nearby LyC-emitting candidates, cosmological simulations, and simulations of LyC escape from star-forming clouds, recent studies are providing insights into the physics of LyC escape and the possible characteristics of the galaxies that reionized the universe. Here, I review progress in LyC detections, the inferred indirect signatures of LyC escape and their application to high redshift, and our current understanding of the physical conditions that lead to high LyC escape. These findings include:

\hangindent=.3cm$\bullet$ LyC-emitting populations are diverse, and multiple factors correlate with LyC escape, particularly neutral gas absorption, dust attenuation, nebular ionization, and concentrated star formation. 

\hangindent=.3cm$\bullet$ Radiative feedback plays a critical role in the youngest starbursts with the highest LyC escape fractions, but mechanical feedback may also contribute. Further research is needed to clarify the timing and role of different feedback mechanisms and to connect local LyC-production sites with the broader interstellar medium. 

\hangindent=.3cm$\bullet$ Indirect LyC diagnostics show promise, but we need to understand whether and how the properties of LyC-emitting galaxies evolve from low to high redshift. 

\end{abstract}

\maketitle

\tableofcontents

\section{INTRODUCTION}
\label{sec:intro}
The reionization of intergalactic hydrogen marks one of the major phase changes of the universe's gas and showcases the impact of early galaxies on their surroundings. After recombination and the formation of the cosmic microwave background, the universe's hydrogen was predominantly neutral. Over time, ultraviolet (UV) photons escaping from early galaxies ionized most of the hydrogen in the intergalactic medium (IGM), a transition known as reionization. \hi\ absorption in quasar and galaxy spectra places the end of reionization around z$\sim$5.3-6, about 1 billion years after the Big Bang, although the exact timeline remains uncertain \citep[\eg][]{fan23, umeda24}. With the recent launch of the {\it James Webb Space Telescope} (JWST) in 2021, we are gaining ever more knowledge about the galaxies that existed during this epoch, as JWST reveals their demographics, morphologies, and stellar and nebular properties \citep[\eg][]{mascia23, endsley23, fujimoto23}. Nevertheless, we have yet to conclusively identify which galaxy populations dominated reionization. 

Photons capable of ionizing hydrogen have energies $E >13.6$ eV ($\lambda < 912$ \AA), a far-ultraviolet wavelength range known as the Lyman continuum (LyC). Both Active Galactic Nuclei (AGN) and short-lived massive O and B stars in star-forming galaxies can produce significant LyC luminosities. The approximate input rate of LyC photons into the IGM is typically represented by the following equation \citep{robertson22}:
\begin{equation}
\label{eqn:reionize}
\dot{n}_{\rm ion} = \rho_{\rm UV} \xi_{\rm ion} f_{\rm esc}^{\rm LyC},
\end{equation}
where $\dot{n}_{\rm ion}$ is the number of ionizing photons per comoving volume per time, $\rho_{\rm UV}$ is the comoving UV luminosity density (erg s$^{-1}$ Hz$^{-1}$ Mpc$^{-3}$) representing the galaxy population, $\xi_{\rm ion}$ is the rate of ionizing, LyC photons produced per UV luminosity (Hz erg$^{-1}$), and \fesc\ is the escape fraction, the fraction of LyC photons that escape galaxies into the IGM. 

Both $\xi_{\rm ion}$ and \fesc\ depend on galaxy properties, and studies continue to debate exactly which galaxy populations dominate the reionization transition. For instance, faint galaxies with absolute UV magnitudes $M_{\rm UV} > -17$ and low \fesc\ ($\lesssim5$ \%) could dominate reionization due to their high $\xi_{\rm ion}$ and greater numbers \citep[\eg][]{finkelstein19, atek24}. This scenario would lead to a more extended reionization timeline with an earlier start and more gradual transition (e.g., 50\%\ of the universe ionized at $z=8.9$; \citealt{finkelstein19}). Alternatively, more luminous galaxies ($M_{\rm UV} < -18$) could contribute almost all the required photons thanks to a higher \fesc\ \citep[\eg][]{naidu20}. The dominance of bright galaxies would lead to a later start to reionization and a more rapid completion (e.g., 50\%\ ionization by $z=6.8$; \citealt{naidu20}). AGN may also have a non-negligible contribution to reionization (see \citealt{robertson22} for a summary), which would reduce the required LyC input from star-forming galaxies \citep[\eg][]{finkelstein19}. Determining \fesc\ and how it depends on galaxy properties is central to resolving these debates. 

While rest-frame UV observations can constrain $\rho_{\rm UV}$, directly detecting escaping LyC from reionization-era galaxies is improbable, if not impossible. Even the LyC light from galaxies at slightly lower redshifts ($z=4-5$) experiences 2-3 magnitudes of attenuation on average as it traverses the cosmic web en route to Earth \citep{inoue14}. As these LyC photons redshift, they are absorbed as LyC or Lyman-series transitions by intervening \hi\ systems. At $z>6$, the LyC light from galaxies must contend with the even more extensive absorbing neutral gas in a partially neutral IGM. 

\begin{marginnote}[] 
\entry{Lyman continuum (LyC)}{Radiation capable of ionizing hydrogen, with $E >13.6$ eV and $\lambda < 912$ \AA.}
\entry{$\rho_{\rm UV}$}{The UV luminosity of galaxies per comoving volume of the universe.}
\entry{$\xi_{\rm ion}$}{The ionizing photon production efficiency, i.e., the number of LyC photons produced per non-ionizing UV luminosity.}
\entry{LyC escape fraction (\fesc)}{The fraction of LyC photons that escape a galaxy and enter the IGM.}
\end{marginnote}

Consequently, many observational studies have focused on constraining \fesc\ using galaxies at $z<4$. At $z\sim3$, LyC redshifts into the visible regime, enabling detections from large ground-based observatories. However, rest-frame optical emission from overlapping lower-redshift galaxies can contaminate the observed LyC signal, leading to false positive LyC detections \citep[\eg][]{vanzella12, siana15}. LyC detections at $z\sim2-4$ therefore require careful confirmation, often by high-resolution imaging. IGM attenuation is also still significant, $\sim1$ magnitude on average at $z\sim3$ \citep{inoue14}, and the precise amount of attenuation along a particular sight line is not known. This uncertainty leads to correspondingly high uncertainties on \fesc\ for individual $z\sim3$ detections \citep[\eg][]{vanzella16} and necessitates stacking larger numbers of galaxies to average over attenuation variations and determine the average \fesc\ for a population \citep[\eg][]{marchi18, steidel18}. Despite these observational challenges, LyC detections in $z=2-4$ galaxies are setting important constraints on \fesc\ and its relationship with galaxy properties at high redshift \citep[\eg][]{vanzella16, riverathorsen17b, marchi18, steidel18, fletcher19, marqueschaves22b}. 

Observations of galaxies at $z\lesssim0.5$ can provide complementary insights into LyC escape. At these low redshifts, IGM attenuation is negligible, which means that studies can explore galaxy-to-galaxy variation in LyC escape without the ambiguity from possible IGM effects. Low redshifts also permit follow-up studies across the electromagnetic spectrum, allowing astronomers to investigate the physical properties of LyC-emitting galaxies from the radio regime to X-rays. Yet, low-redshift LyC observations are not without their drawbacks. Typical low-redshift star-forming galaxies likely contribute little LyC emission to the IGM, with the $z\sim0$ extragalactic ionizing background instead dominated by AGN \citep[\eg][]{fauchergiguere20}. In order to study LyC escape under conditions relevant at high redshift, low-redshift targets must resemble high-redshift galaxies in at least some properties. Nevertheless, low-redshift galaxies need not be perfect analogs to be useful, since they can still demonstrate which conditions do and do not affect \fesc\ and why. 

In this review, I focus on the physical insights we are gaining from studies of LyC escape at $z\lesssim0.5$ as well as the open questions that remain. Over the past decade, the number of LyC detections at $z\lesssim0.5$ has grown from a handful of galaxies to large samples, enabling quantitative estimates of \fesc\ and its variation among galaxy populations. Multi-wavelength studies are revealing the physical and observable properties that characterize these confirmed LyC emitters (LCEs). Along with detailed studies of similar galaxies at even closer distances, these observations are helping us understand the physical mechanisms and interstellar gas geometries responsible for LyC escape. Observations of $z\lesssim0.5$ LCEs have also led to the development of new, indirect diagnostics for predicting \fesc, which are now being applied to JWST observations of galaxies in the epoch of reionization. 

\begin{marginnote}[] 
\entry{LCE}{Lyman continuum emitter}
\end{marginnote}

A comprehensive understanding of reionization and the physics of LyC escape requires information from a variety of subfields, including research on the IGM, cosmological simulations, AGN, stellar populations, and high-redshift galaxies. This review covers $z\lesssim0.5$ observations of star-forming galaxies and related theoretical results. For a recent review on reionization, including LyC observations at higher redshift, see \citet{robertson22}. The recent review by \citet{fan23} summarizes reionization timing constraints from quasar observations. Lastly, I focus here on the escape of ionizing photons, but constraining $\xi_{\rm ion}$ and understanding the production of ionizing photons is equally important. Reviews by \citet{robertson22} and \citet{eldridge22} summarize recent constraints on $\xi_{\rm ion}$ and the state of our knowledge about ionizing photon production from massive stars.

\section{LYC MEASUREMENTS AND METHODS}
\label{sec:methods}
\subsection{Calculating \fesc}
\label{sec:methods:direct}
Observationally constraining \fesc\ requires comparing the flux of escaping LyC with an estimate of the intrinsic LyC produced in a galaxy. At low redshift, LyC measurements probe wavelengths near the Lyman edge (typically $\sim 880-900$ \AA, but as low as 820 \AA), an observational constraint set by the wavelength-dependent sensitivity of UV telescopes. Absorption by Milky Way hydrogen sets a lower redshift limit for LyC observations of $z\sim0.01$, but such low redshift observations require careful masking of geocoronal Lyman series emission lines \citep[\eg][]{leitherer16}. To date, most low-redshift LyC detections have been made at $z\sim0.3$ using the Cosmic Origins Spectrograph (COS) on the {\it Hubble Space Telescope} (HST). The COS sensitivity drops sharply below $\lambda \sim1150$ \AA\ \citep{hirschauer23}, corresponding to the redshifted Lyman limit for a galaxy at $z=0.26$. LyC observations at $z < 0.26$ are therefore extremely challenging. 

Estimating the intrinsic LyC in order to derive \fesc\ is not straightforward. One method involves fitting stellar population synthesis models to the non-ionizing UV spectra and obtaining the intrinsic LyC flux from the best-fit spectrum \citep[\eg][]{saldanalopez22}. Uncertainties in stellar population properties, model degeneracies, and uncertainties in the far-UV spectra of massive stars can affect the accuracy of the resulting \fesc\ estimates. Despite the challenges in detecting the LyC, the uncertainty in estimating the intrinsic LyC produced typically dominates the uncertainty in \fesc\ \citep{saldanalopez22}. An alternative method uses nebular Balmer emission lines, typically H$\beta$, to estimate the total number of LyC photons absorbed by nebular gas \citep[\eg][]{leitherer95, izotov16a, flury22a}. The \fesc\ is then given by
\begin{equation}
\label{eqn:feschb}
f_{\rm esc}({\rm H}\beta) = \frac{F_{\lambda {\rm LyC}}^{obs}}{F_{\lambda {\rm LyC}}^{obs}+F_{\lambda {\rm LyC}}^{abs}},
\end{equation}
where $F_{\lambda {\rm LyC}}^{obs}$ is the observed LyC and $F_{\lambda {\rm LyC}}^{abs}$ is the absorbed LyC flux calculated from H$\beta$ \citep{izotov16b, flury22a}. A related method fits spectral energy distribution (SED) models, including Balmer line and nebular emission, to the optical spectrum \citep[\eg][]{izotov16b}. Although H$\beta$-derived estimates of the absorbed LyC serve as a useful check on estimates from UV SED fits, the H$\beta$ method has its own disadvantages. It still requires information about the stellar population, because the conversion between 900 \AA\ flux and H$\beta$ flux depends on the ionizing spectral shape \citep[\eg][]{izotov16b, flury22a}. Different stellar population models and adopted star formation histories can lead to a factor of 2-3 difference in \fesc, especially for older starbursts \citep{flury22a}. H$\beta$ also only traces LyC absorption by gas and neglects LyC absorption by dust, which can dominate the LyC opacity in some objects \citep[\eg][]{borthakur14}. More problematically, the H$\beta$ method assumes isotropic escape. While the observed LyC ($F_{\lambda {\rm LyC}}^{obs}$) is measured along the line of sight, the H$\beta$-emitting gas may have been ionized by LyC photons from other directions. LyC escape is likely anisotropic \citep[\eg][]{cen15}, but Equation~\ref{eqn:feschb} assumes that the observed and absorbed LyC are the same in all directions. 

Instead of the absolute \fesc, some studies report alternative measures of the escaping LyC. The relative escape fraction, $f_{\rm esc,rel}$, compares the observed ratio of ionizing and nonionizing fluxes to the intrinsic ratio predicted from stellar population models \citep[\eg][]{leitet13, steidel18, wang19}. One common definition is
\begin{equation}
\label{eqn:fescrel}
f_{\rm esc,rel} = \frac{(f_{\rm 1500}/f_{\rm 900})_{\rm int}}{(f_{\rm 1500}/f_{\rm 900})_{\rm obs}} = f_{\rm esc} \times 10^{0.4\times A_{\rm1500}},
\end{equation}
which compares the observed and intrinsic 900 to 1500 \AA\ flux ratios and which is related to the absolute escape fraction by correcting the observed 1500 \AA\ flux for the internal dust attenuation ($A_{\rm 1500}$). However, the exact definition of $f_{\rm esc,rel}$, including the treatment of dust, varies in the literature (see \citealt{steidel18} for a discussion). In general, $f_{\rm esc,rel}$ is higher than $f_{\rm esc}^{\rm LyC}$ and predominantly traces the absorption of LyC by HI and not dust. Other works consider purely observational quantities, such as the observed ionizing to non-ionizing flux ratios ($f_{\rm 900}/f_{\rm 1100}$ or $f_{\rm 900}/f_{\rm 1500}$; e.g., \citealt{flury22a}) or the observed LyC flux. When converted to luminosity, the latter quantity is ultimately what matters for the IGM ionization budget, and the sources with the highest LyC luminosities may not necessarily have the highest \fesc\ \citep[\eg][]{komarova24}. 

\begin{marginnote}[]
\entry{Relative escape fraction (\fescrel)}{A comparison of the observed versus intrinsic ionizing to nonionizing flux ratios.}
\end{marginnote}

The fraction of $\sim900$ \AA\ photons that escape may differ from the total fraction of LyC photons that escape. Near the Lyman edge, the cross-section for LyC absorption by hydrogen depends on wavelength approximately as $\sigma \propto \lambda^3$ \citep[\eg]{osterbrock06}. Because of this dependence, the $f_{\rm esc}$ at 900 \AA\ would underestimate the total \fesc, and even galaxies with near-zero $f_{\rm esc}$ at 900 \AA\ (<1\%) could have a total \fesc\ as high as 10-20\%\ \citep{mccandliss17}. The wavelength dependence of dust attenuation may also affect the conversion between the 900 \AA\ and total \fesc\ \citep[\eg][]{leitet13, mccandliss17} but is poorly constrained below the Lyman limit. 

The possible contribution of nebular LyC represents a further complication. Recombinations to the hydrogen ground state generate LyC photons. Since lower energy electrons have a higher probability of recombining, nebular LyC peaks at longer wavelengths than the intrinsic stellar LyC \citep{inoue10}. Nebular LyC may therefore increase the ionizing flux toward longer wavelengths or even produce an emission bump near the Lyman limit \citep{inoue10, simmonds24b}. The observed $f_{\rm esc}$ at 900 \AA\ would then be artificially high, by up to a factor of two, compared to the stellar $f_{\rm esc}$ at 900 \AA, although it may coincidentally more closely approximate the total stellar \fesc\ across all wavelengths \citep{simmonds24b}. The effect of nebular LyC also depends critically on the gas optical depth and structure. Nebular LyC increases the 900 \AA\ signal most strongly at moderately opaque hydrogen column densities ($N_{\rm HI} \sim 10^{17}-10^{18}$ cm$^{-2}$; \citealt{simmonds24b}). However, the effect of nebular LyC may not be as strong for a clumpy medium, where the optically thick, nebular LyC-producing regions are distinct from the transparent regions through which LyC escapes \citep{flury24}. Existing $z\sim0.3$ observations do not show increased flux near the Lyman limit \citep{flury24} but only cover a limited wavelength range (typically 870-890 \AA). Future observations and simulations of realistic gas distributions are needed to establish the spectral shape of the LyC and its effect on the inferred \fesc.

\subsection{Indirect Constraints on \fesc}
\label{sec:methods:indirect}
Because of the limited sensitivity of COS and the {\it Far Ultraviolet Spectroscopic Explorer} (FUSE) at short wavelengths, directly detecting LyC has not been possible for most galaxies at $z<0.3$. Nevertheless, galaxies at the lowest redshifts can provide important information about LyC escape, allowing us to study stellar feedback, the interstellar medium (ISM) geometry, and the propagation of radiation at high spatial resolution. Studies of LyC escape in these nearby galaxies typically rely on indirect estimates of optical depth. 

\subsubsection{Idealized Geometries}
\label{sec:methods:indirect:geometry}
\begin{figure}[h]
\centering
\begin{subfigure}[b]{0.5\linewidth}
\includegraphics[width=\linewidth]{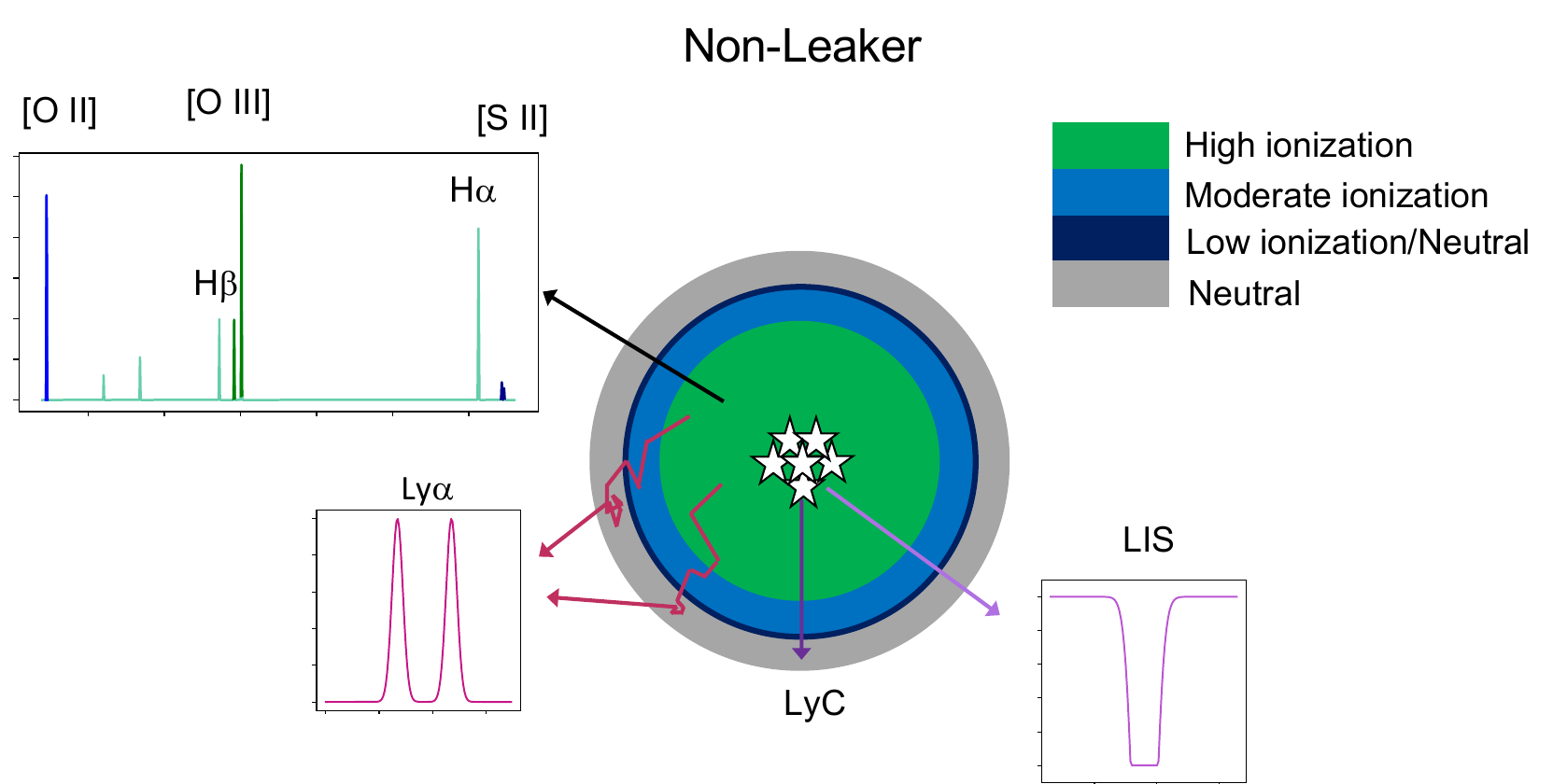}
\end{subfigure}

\begin{subfigure}[b]{0.5\linewidth}
\includegraphics[width=\linewidth]{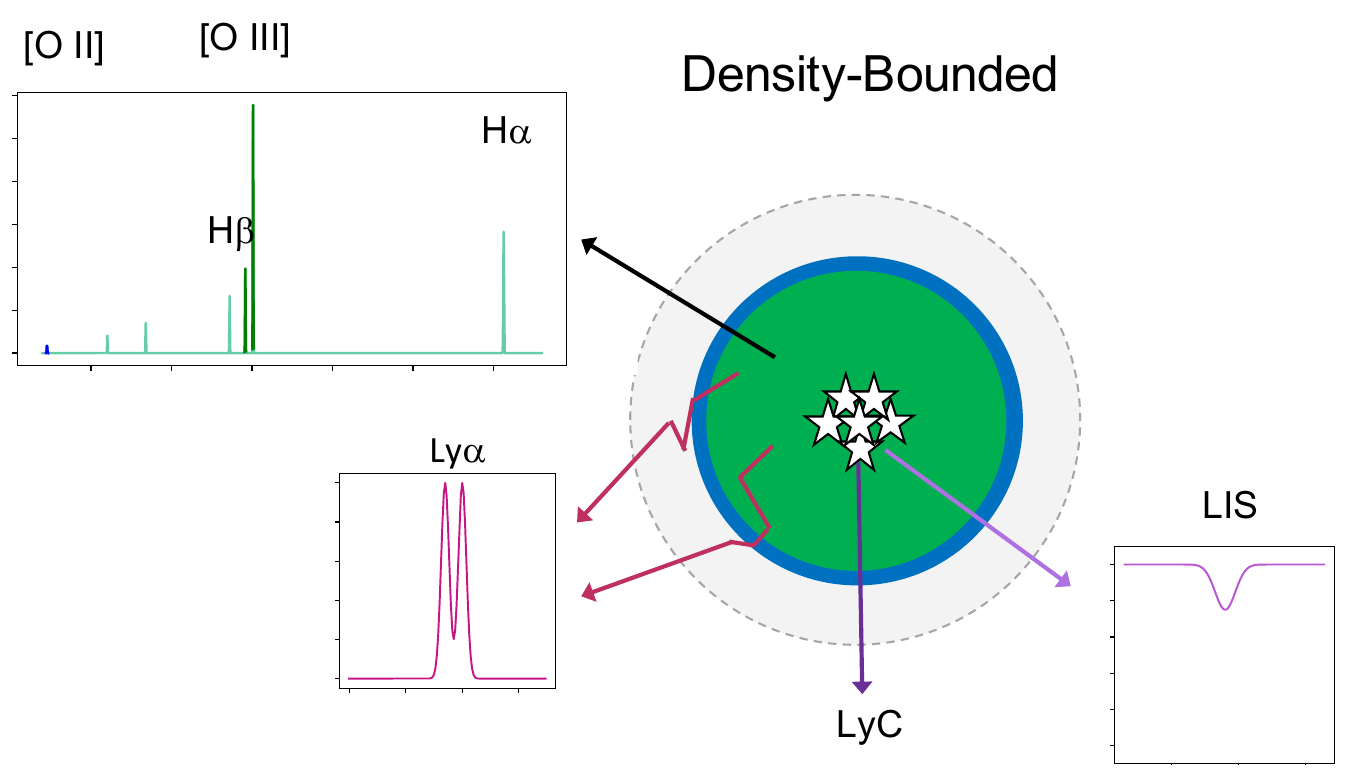}
\end{subfigure}
\begin{subfigure}[b]{0.5\linewidth}
\includegraphics[width=\linewidth]{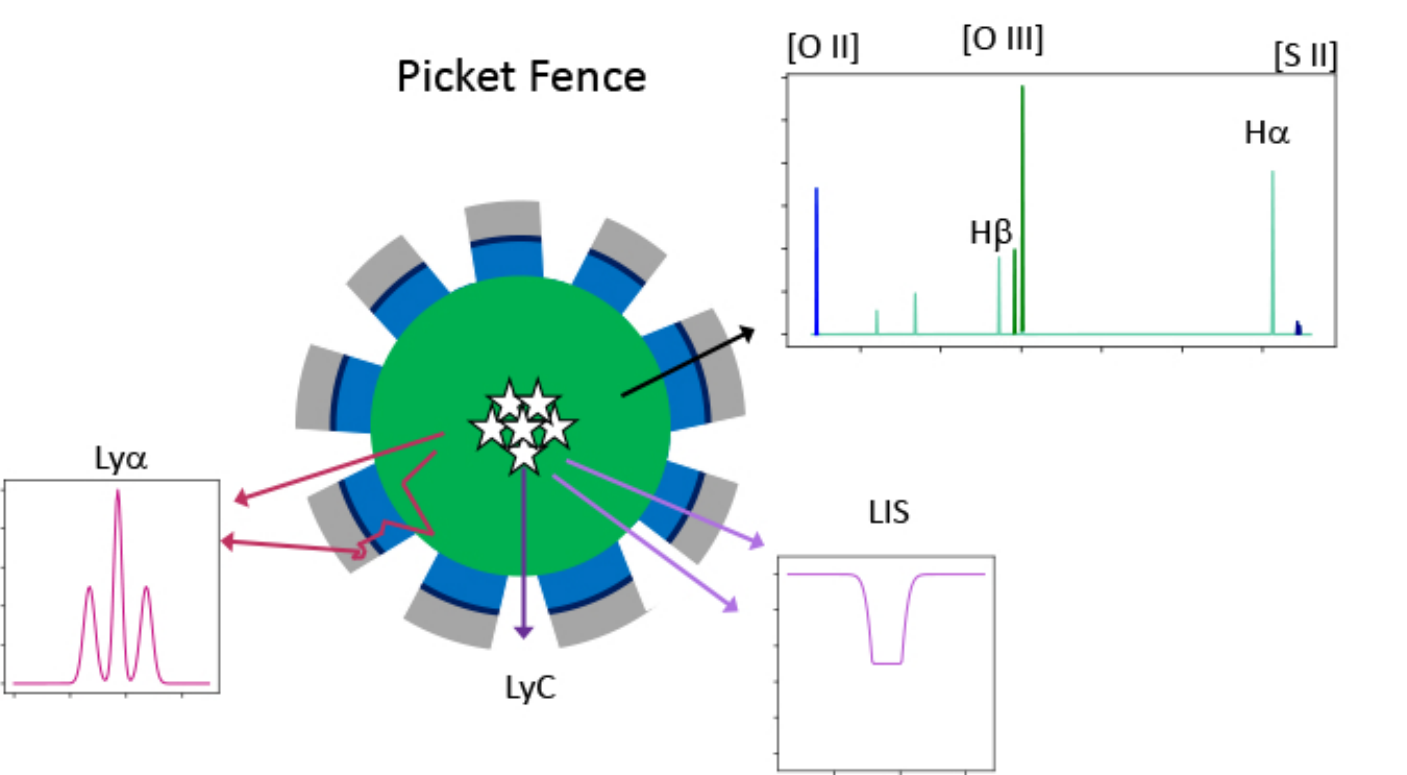}
\end{subfigure}
%
\caption{Idealized models of LyC escape. Non-leaking geometries (top) are classic Str{\"o}mgren spheres, with nebular emission from low- and high-ionization gas. Nebular \lya\ emission scatters on neutral H as it propagates, resulting in a broad, double-peaked emission profile. LIS metals absorb UV photons to form deep absorption lines reaching to zero flux. Gas in density-bounded models (center) does not extend to the Str{\"o}mgren radius and low-ionization emission and absorption lines weaken. \lya\ scatters on trace amounts of neutral gas, resulting in a narrower emission profile. In the picket fence model (bottom), some sight lines are transparent, while others are opaque to the LyC. The optical spectrum shows a mix of low- and high-ionization emission lines. The sight lines along which LyC photons escape show high ionization. If the \hi\ column is low enough, \lya\ photons can be transmitted directly through holes, resulting in an emission peak at line center. LIS absorption lines are saturated, but the non-zero flux at the absorption line center indicates a partial covering fraction of low-ionization gas. The \lya\ profiles shown above are idealized and neglect dust and kinematics, which can also affect the line shape \citep[\eg][]{verhamme15}. }
\label{fig:diagnostics}
\end{figure}

Many discussions regarding LyC escape and proposed indirect diagnostics emphasize two simple models: a ``density-bounded" model and the ``picket fence" model (Figure \ref{fig:diagnostics}). A ``radiation-" or ``ionization-bounded" nebula represents the classic Str{\"o}mgren sphere, where gas fully absorbs all ionizing radiation and the balance between recombination and ionization rates sets the \hii\ region radius. In contrast, a density-bounded nebula leaks LyC radiation, as the surrounding gas is insufficient to fully absorb the available ionizing radiation. Here, the extent of the absorbing gas sets the size of the ionized nebula. In the simplest case, a density-bounded nebula is a homogeneous sphere, where the \hi\ column density in all directions is $N_{\rm HI}\lesssim10^{17}$ cm$^{-2}$ and $N_{\rm HI}$ determines \fesc. 

The picket fence model presents a contrasting picture of a LyC-leaking region. In this model, dense clumps of gas fully absorb LyC in some directions, and LyC escapes along transparent channels between these clumps \citep{heckman01}. In this case, the covering fraction (\cf) of optically thick gas determines \fesc. In the picket fence model, dust may be associated only with the dense clumps, in which case it does not affect \fesc, or it may be modeled as a uniform screen, in which case it will reduce the number of escaping photons \citep[\eg][]{gazagnes18}. Of course, the true ISM geometry of a galaxy is likely more complex than either the simplistic density-bounded or picket fence scenarios \citep[\eg][]{jaskot19}. In the most extreme LCEs, the absorbing clumps may not be fully opaque, thereby allowing partial LyC escape in some directions \citep{gazagnes20}, and the distribution of \nhi\ in simulated galaxies is more complex than the picket fence model \citep{mauerhofer21}. Nevertheless, simple models can help build intuition regarding possible observable signatures of LyC escape.

\begin{marginnote}[] 
\entry{Density-bounded}{A nebula with insufficient gas to absorb all ionizing radiation and a radius less than the Str{\"o}mgren radius.}
\entry{Picket fence}{A simple model where LyC escapes through transparent channels in between optically thick gas.}
\entry{Covering fraction (\cf)}{The fraction of optically thick gas that blocks the line of sight to an emitting source.}
\end{marginnote}

\subsubsection{Indirect Signs of LyC Escape}
\label{sec:methods:indirect:signs}
Density-bounded nebulae should lack emission from the outer edge of the Str{\"o}mgren sphere (Figure \ref{fig:diagnostics}), corresponding to a deficit in lower ionization emission lines, such as \oii~$\lambda$3727, \sii~$\lambda\lambda$6717,6731, and \oi~$\lambda$6300 \citep[\eg][]{mccall85, iglesiasparamo02, giammanco05, pellegrini12}. The technique of ionization parameter mapping (IPM) traces the possible paths of LyC escape by mapping the ratio of high- to low-ionization lines, such as \oiii~$\lambda$5007/\sii~$\lambda\lambda6717,6731$ or \oiii~$\lambda$5007/\oii~$\lambda3727$ \citep{pellegrini12}. Candidate sites for LyC escape should remain highly ionized in a given direction, without transitioning to lower-ionization emission. Unlike direct LyC measurements, IPM can potentially identify LyC escaping transverse to the line of sight. This mapping technique has been applied to \hii\ regions in the LMC, SMC, and Haro 11, as well as other local starburst galaxies (Section \ref{sec:local}; \citealt[\eg][]{pellegrini12, zastrow13, keenan17, bik18}). Line ratios do not unambiguously identify LyC-leaking regions, however, as they are affected by factors such as dust attenuation, ionization parameter, and the ionizing source spectrum \citep[\eg][]{giammanco05, pellegrini12}. 

\begin{marginnote}[]
\entry{Ionization Parameter Mapping (IPM)}{A technique that identifies possible optically thin regions via their high nebular ionization ratios. }
\end{marginnote}

Low optical depths along the line of sight should also leave a signature in the UV spectra of galaxies (Figure \ref{fig:diagnostics}). With an optical depth $\sim10^4$ times greater than that of the LyC \citep{verhamme15}, the Lyman-$\alpha$ (\lya) emission line is easily absorbed by neutral hydrogen. Unlike LyC, \lya\ resonantly scatters, with absorption followed by re-emission in a random direction. Lower \hi\ optical depths should thus lead to less resonant scattering and more \lya\ escape closer to the line center. \lya\ emission with a peak at the systemic velocity may indicate escape through a transparent hole, while low-column density shells of gas or porous, clumpy gas can produce narrow, double-peaked \lya\ emission lines \citep{behrens14, verhamme15, dijkstra16}. Other studies suggest that strong \lya\ flux blueward of the systemic velocity may indicate low optical depth \citep{heckman11} or that the asymmetry of the red \lya\ peak may trace the ISM porosity \citep{kakiichi21}. Nevertheless, \lya\ radiative transfer is a complicated phenomenon, and \lya\ photons can scatter into the line of sight from other directions. As an indirect diagnostic, \lya\ can identify plausible LCE candidates but may not guarantee LyC escape \citep[\eg][]{flury22b, schaerer22}. 

\begin{marginnote}[]
\entry{Low-ionization state (LIS)}{Ions with ionization potentials $<13.6$ eV, which can co-exist with neutral hydrogen gas.}
\end{marginnote}

A deficiency in neutral gas along the line of sight will also lead to weak absorption features from the neutral ISM (Figure \ref{fig:diagnostics}). Lyman series lines in the far UV (FUV) may not extend to the zero flux level \citep{gazagnes18}, and low-ionization state (LIS) metal absorption lines from \cii, \sitwo, or \oip\ should also be weak. Saturated absorption lines showing net residual flux at the line center may imply a low \cf, where the absorption represents the portion of dense clouds that cover the UV-emitting source \citep{heckman01, heckman11}. However, the observed absorption line depths will also depend on spectral resolution, gas metallicity, and kinematics \citep[\eg][]{reddy16b, gazagnes18, saldanalopez22}, and resonant emission can infill some of the absorption \citep[\eg][]{prochaska11}. Wavelength-dependent dust extinction and differences in the UV SEDs of leaking and non-leaking populations can also affect the relationship between \fesc\ and LIS line depths \citep{leitet13}. Both \lya\ emission and LIS absorption lines have been used to infer low optical depths in galaxies without direct LyC measurements (Section \ref{sec:local}; e.g., \citealt{heckman11, leitet13, verhamme15, henry15, jaskot17, izotov20}), and all these indirect indicators (IPM, \lya\, and LIS lines) have some support from confirmed LyC detections (Section \ref{sec:lces}; \citealt{izotov18b, flury22b}). 

Finally, for the closest galaxies, with resolved stellar populations, studies can estimate \fesc\ by comparing the expected ionizing photon production with the LyC absorption traced by dust and ionized gas emission \citep[\eg][]{oey97, choi20}. If the dust and gas do not fully account for the ionizing photon supply, the missing ionizing photons have presumably escaped. With this technique, one can estimate the global, not the line-of-sight, \fesc\ from individual \hii\ regions and from entire galaxies. However, the exact values of \fesc\ are sensitive to the adopted stellar population models, with higher \fesc\ inferred for stellar models with higher LyC production \citep{oey97, voges08}.

\section{HISTORY OF LYC MEASUREMENTS AT LOW REDSHIFT}
\label{sec:history}
\subsection{Early LyC Measurements in Nearby Starbursts}
\label{sec:history:early}
\begin{figure}[h]
\centering

\begin{subfigure}[b]{1\textwidth}
\caption{}
\includegraphics[width=\linewidth]{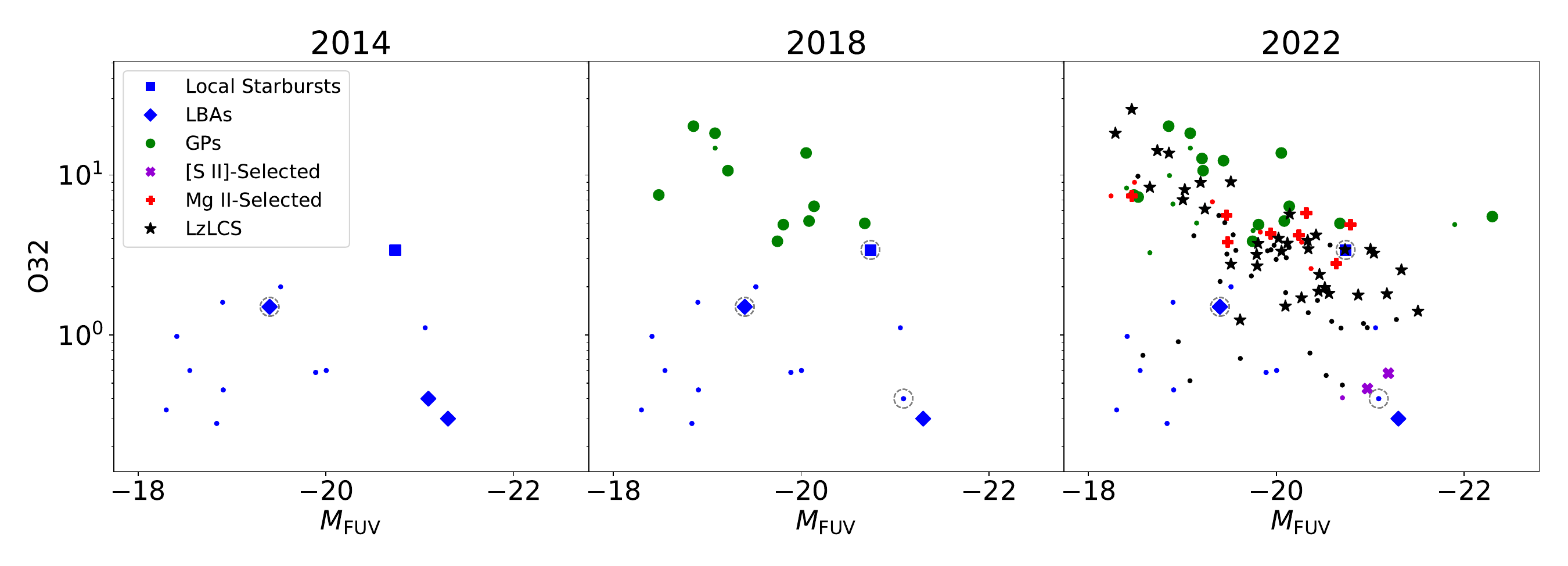}
\end{subfigure}

\begin{subfigure}[b]{1\textwidth}
\caption{}
\includegraphics[width=\linewidth]{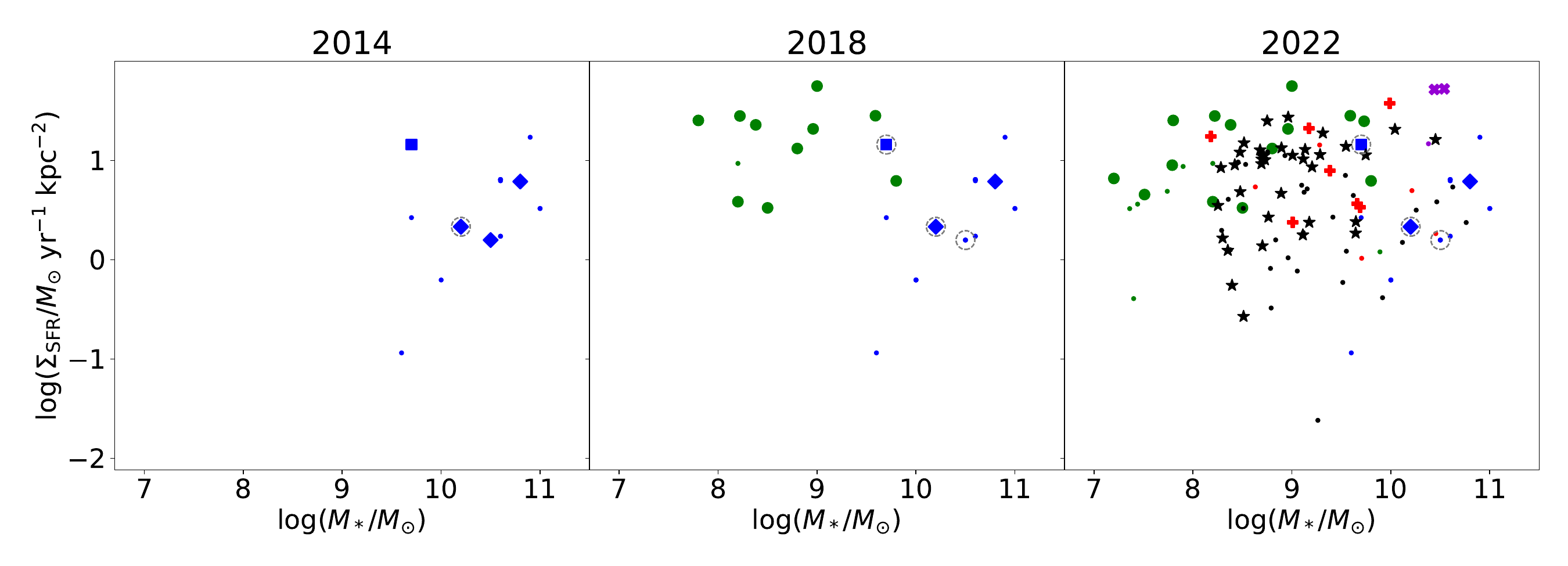}
\end{subfigure}

\caption{The parameter space in O32 and observed $M_{\rm FUV}$ (Panel a) and in \sigsfr, and $M_*$ (Panel b) covered by LyC observations from 2014-2022. Symbols indicate different sample selections. Large symbols show LCEs, and dots show non-detections. Gray, dashed circles mark debated detections. Early LyC observations were limited to massive, low-ionization starbursts. By 2018, many lower mass, highly ionized GPs were known LCEs. As of 2022, LyC samples span a wide range of parameters and selection methods. Galaxy properties and samples come from \citet{calzetti92, vacca92, leitherer95, leitherer02, leitherer16, calzetti00, deharveng01, ostlin01, kong02, windhorst02, keel05, bergvall06, grimes06, grimes07, grimes09,  lopezsanchez06, moustakas06, leitet11, leitet13, overzier11, atek14a, borthakur14, alexandroff15, chisholm15, chisholm17, chisholm18, heckman15, izotov16a, izotov16b, izotov18a, izotov18b, izotov21, izotov22,  puschnig17, verhamme17, wang19, malkan21, flury22a, xu22, melinder23}, and archival COS acquisition images. $M_*$ for the \citet{xu22} sample are courtesy of X. Xu. When possible, \sigsfr\ uses the UV half-light radius, and $M_{\rm FUV}$ corresponds to $\sim1500$ \AA; in a few cases, I use optical radii and the 1150 or 1600 \AA\ magnitude. (See Supplemental Animations 1 and 2 for animated figures covering the years 1995-2022.)}
\label{fig:history}
\end{figure}

Early efforts to detect LyC focused on nearby galaxies at $z\sim0.01-0.04$ ($\sim70-200$ Mpc) and found evidence for little, if any, LyC escape. Figure \ref{fig:history} summarizes LyC observations and detections from 1995 to 2022. \citet{leitherer95} analyzed LyC observations from the {\it Hopkins Ultraviolet Telescope} for two nuclear starbursts and two blue compact dwarf galaxies and placed upper limits of 0.95-15\%\ on \fesc. FUSE observations of eight other galaxies at similar redshift also resulted in non-detections, with \fesc\  $< 0.2-16$\%\ \citep{grimes09, leitet13}, as well as three debatable detections described below. 

The nearby galaxies Haro 11, Mrk 54, and Tol 1247-232 have been targeted by both FUSE and HST/COS but with conflicting results. \citet{bergvall06} claimed the first LyC detection in Haro 11, \fesc$=4-10$\%, with FUSE. \citet{grimes07, grimes09} disputed this detection, arguing it resulted from uncertainties in the background subtraction and geocoronal contamination, and placed upper limits of \fesc\ $< 2$\% and $<1.6$\%. A re-analysis of the FUSE data by \citet{leitet11} addressed these issues and confirmed LyC escape, but with a lower \fesc $= 3.3$\%. Recent HST/COS observations detect LyC from two of Haro 11's three knots, with a total \fesc$=3.9 \pm 3.4$\%\ \citep{komarova24}; these high uncertainties are consistent with most of the prior detections and limits. For Mrk 54, an upper limit of \fesc$<6.2$\%\ from FUSE \citep{deharveng01} was followed by a detection with \fesc$=2.5$\% from HST/COS \citep{leitherer16}. However, \citet{chisholm17} argue that the observed LyC could be geocoronal emission, because observations taken within Earth's shadow suggest a non-detection. Similarly, \citet{leitet13} derive \fesc$=2.4$\%\ for Tol 1247-232 from FUSE data, whereas analyses of COS observations report \fesc$=0.4$\%, 1.5\%, and 4.5\%, depending on the treatment of the dark current and geocoronal emission \citep{chisholm17, puschnig17, leitherer16}. These disagreements highlight the difficulty in distinguishing the weak LyC emission from confounding signal from dark current, background, and geocoronal emission. Regardless, FUSE and COS observations have established that the \fesc\ from many low-redshift star-forming galaxies is low, only a few percent at most.

To explain the dearth of LyC detections, \citet{bergvall13} suggested that starburst selection methods were to blame. Selecting on high H$\alpha$ equivalent widths (EWs) identifies galaxies with a strong burst of recent star formation but also inherently selects galaxies with substantial ionized gas from the absorption of LyC photons. \citet{bergvall13} proposed instead selecting LCE candidates on blue UV colors and weak nebular emission lines. However, like other proposed indirect diagnostics, this method is not free of degeneracies, as the relationship between the UV spectral slope $\beta$, H$\beta$ EW, and \fesc\ depends on factors such as dust and the stellar SED \citep[\eg][]{zackrisson13}. For instance, an aging starburst could also present blue colors and weak nebular emission, mimicking a low optical depth \citep{zackrisson13}.

By virtue of their similarities to high-redshift Lyman Break Galaxies (LBGs), low-redshift Lyman Break Analogs (LBAs) represent a population of proposed LCE candidates. LBAs share properties such as UV luminosity, stellar mass, metallicity, morphology, and dust content with LBGs, and this class includes the possible LCEs Haro 11 and Mrk 54. \citet{heckman11} examined LIS absorption lines from FUSE and COS spectra of 11 LBAs and 15 UV-bright galaxies at redshifts $z=0.001-0.23$, finding evidence of low \cf\ in four galaxies. They argued that three of these galaxies were likely LCEs, based on supporting evidence from blue-shifted \lya\ emission and low H$\alpha$/UV flux ratios suggesting reduced ionized gas emission. These same galaxies exhibit compact massive star-forming regions in their centers, which drive outflows up to 1500 km s$^{-1}$. Follow-up COS observations of one candidate LCE LBA confirmed LyC escape with \fescrel$ = 21$\%, consistent with the LIS line constraints \citep{borthakur14}. However, because of significant dust absorption in this galaxy, the absolute \fesc\ is much lower, only 1\%. 

\begin{marginnote}[] 
\entry{Lyman Break Galaxy (LBG)}{A high-redshift (typically $z>3$) galaxy, selected by the apparent presence of the Lyman break spectral feature in photometric filters.}
\entry{Lyman Break Analog (LBA)}{Low-redshift galaxies selected to resemble LBGs in their UV luminosities and other properties.}
\end{marginnote}

These few LyC detections, non-detections, and indirect LIS line estimates offered some clues as to which factors might correlate with LyC escape. Using direct LyC measurements and indirect estimates from \cii~$\lambda1036$ absorption, \citet{leitet13} suggested that \fesc\ may increase with lower stellar mass, higher specific SFR (sSFR), and lower dust attenuation. \citet{alexandroff15} examined three potential indirect signs of LyC escape in LBAs (LIS lines, blue-shifted \lya, and weak \sii\ emission) and found that the most promising LCE candidates were the LBAs with high SFRs per unit area (\sigsfr) and fast outflows. This result suggested that supernova (SN) feedback and intense radiation from a concentrated burst of star formation could clear out gas and enhance LyC escape \citep{heckman11, alexandroff15}. %

\subsection{The Green Peas}
\label{sec:history:gps}
As discussed in Section \ref{sec:methods:indirect:signs}, a deficit in low-ionization emission could potentially identify optically thin \hii\ regions. Discovered by citizen scientists in the Sloan Digital Sky Survey (SDSS) Galaxy Zoo project, Green Pea (GP) galaxies are extreme emission line galaxies whose green color results from strong \oiii~$\lambda$5007 and comparatively weaker \oii~$\lambda3727$ emission \citep{cardamone09}. These galaxies share many properties with high-redshift \lya\ emitters (LAEs), such as high nebular ionization, compact sizes, high sSFRs, and strong \lya\ \citep{nakajima14, henry15, yang17b, kim21}. Based on their high \oiii~$\lambda$5007/\oii~$\lambda$3727 (O32) ratios, \citet{jaskot13} and \citet{nakajima14} proposed that the GPs could be density-bounded LCEs, although photoionization models were inconclusive. \citet{stasinska15} demonstrated that the galaxies with the highest O32 also had the highest H$\beta$ EWs, which argued against LyC escape and in favor of high ionization parameters as the origin of high O32.

\begin{marginnote}[]
\entry{Green Pea (GP)}{A compact starburst galaxy, identified by strong nebular \oiii~$\lambda$5007 emission.}
\end{marginnote}

A series of papers by Izotov et al.\ conclusively settled the question of LyC escape in the GPs. Using HST/COS observations of 11 GPs at an optimal redshift range of $z=0.29-0.43$, \citet{izotov16a, izotov16b, izotov18a, izotov18b} detected LyC emission in all targeted GPs with \fesc$=2-72$\%, with the intrinsic LyC estimated from H$\beta$. These detections marked a clear departure from the weak to absent LyC in prior samples and demonstrated that the GPs represent a population with LyC escape as a nearly ubiquitous characteristic. 

The GPs provided new information about which galaxy properties might be related to LyC escape. Although O32 $>5$ effectively selected LCEs, the relationship between O32 and \fesc\ showed high scatter \citep{izotov18b}.  The GPs' UV spectra also supported the hypothesized links between \lya\ emission, weak LIS lines, and LyC escape. The LCE GPs show strong \lya\ emission, with the \lya\ escape fraction (\fesclya) typically greater than or equal to the LyC \fesc\ \citep{verhamme17}. A narrow velocity separation of the red and blue \lya\ peaks ($v_{\rm sep}$) correlates tightly with the observed \fesc, and the strongest LCE GP exhibits a \lya\ peak near the line center, as expected for transmission through a transparent hole \citep{verhamme17, izotov18b}. Based on the residual fluxes in the Lyman series absorption lines and the inferred UV dust attenuation, \citet{chisholm18} and \citet{gazagnes20} show that the combination of low \cf\ and low dust extinction can explain the GPs' observed \fesc. The LCE GPs also have high ionization parameters, low metallicities ($12+\log_{\rm 10}(O/H) \lesssim8.2$), low stellar masses (\logten($M_*$/\Msol)$=7.8-9.8$), and high \sigsfr, properties which could also be associated with high \fesc\ \citep{verhamme17,izotov18b,chisholm18}. Because of the GPs' young inferred starburst ages ($\lesssim 3$ Myr), \citet{izotov18b} suggest that stellar winds and radiation, rather than SN feedback, could be responsible for their high \fesc. Such young ages also imply high $\xi_{\rm ion}$ and thus high LyC luminosities \citep{izotov16a}. 

\subsection{Additional Nebular Emission Selected LCEs}
\label{sec:history:nebular}
Following the detection of LyC in the GPs, other works have investigated nebular-line selected LCE candidates. \citet{izotov21} measured \fesc\ in nine lower-mass (\logten($M_*$/\Msol)$ < 8.0$) GPs with O32$=4-14$ but detected LyC in only four galaxies. This result shows that \fesc\ does not increase with lower $M_*$ for O32 or EW-selected galaxies, but \fesc\ remains unconstrained for low-mass galaxy samples with O32 outside of the investigated range. \citet{malkan21} report \fesc$\sim4$\%\ and an upper limit of \fesc$<1$\%\ for two more luminous GPs with absolute rest-frame FUV magnitudes $M_{\rm FUV} \sim -22$ at $z=0.5$. \citet{wang19} investigated an alternative nebular selection method, choosing three LCE candidates based on an offset to lower \sii~$\lambda\lambda6717,6731$/H$\alpha$ ratios in an \oiii~$\lambda$5007/H$\beta$ versus \sii/H$\alpha$ diagram. They detected LyC escape with \fesc$=2-4$\% in the two more compact galaxies. Unlike GPs, these new LCEs were more massive (\logten($M_*$/\Msol)$=10.4-10.5$), more metal-rich ($12+\log_{\rm 10}(O/H) =8.6-8.7$), less highly ionized (O32 $<2$), and weaker in \lya\ emission. They also show high dust content, which results in a lower absolute \fesc\ in comparison with their much higher \fescrel$=80-93$\%. Finally, \citet{izotov22} and \citet{xu22} find that the resonant \mgii~$\lambda\lambda2796,2803$ doublet can select LCE candidates, with \fesc$=0.7-13.6$\%\ in the confirmed LCEs.

\subsection{The Low-redshift Lyman Continuum Survey}
\label{sec:history:lzlcs}
The growing number of low-redshift LyC detections pointed the way to possible LCE selection criteria but still suffered from relatively small numbers. The Low-redshift Lyman Continuum Survey (LzLCS), a large program with HST/COS, expanded on these efforts by targeting 66 additional galaxies in the LyC to systematically test proposed diagnostics of LyC escape \citep{flury22a}. LzLCS galaxies were selected on high O32, high \sigsfr, or blue UV slopes $\beta$, with the goal of evenly selecting galaxies that met one, two, or all three of these criteria. As intended, the LzLCS galaxies spanned a wider range of metallicities, O32, EWs, and sizes compared with previous selections. \citet{flury22a} reprocessed the COS spectra and remeasured LyC for literature samples \citep{izotov16a, izotov16b, izotov18a, izotov18b, izotov21, wang19}, calculating \fesc\ from the UV SED, \fesc\ from H$\beta$, and the \fratio\ ratio for a combined sample of 89 $z\sim0.3$ galaxies, including 50 LyC detections. Although not included in this combined LzLCS sample, the \citet{izotov22} and \citet{xu22} samples bring the total number of $z\sim0.3$ LyC detections to 58 and non-detections to 45. While trends using different \fesc\ estimates generally agree, the SED-based estimates have the advantage of not assuming isotropy (Section~\ref{sec:methods}), and I adopt these values unless otherwise specified. The combined LzLCS sample (LzLCS+) set the stage for statistical investigations of \fesc\ and its relation to galaxies' physical and observable properties. 

\section{PROPERTIES OF LYC EMITTERS}
\label{sec:lces}
\begin{figure*}[h]
\centering
\begin{subfigure}[b]{0.64\linewidth}
\caption{}
\includegraphics[width=\linewidth]{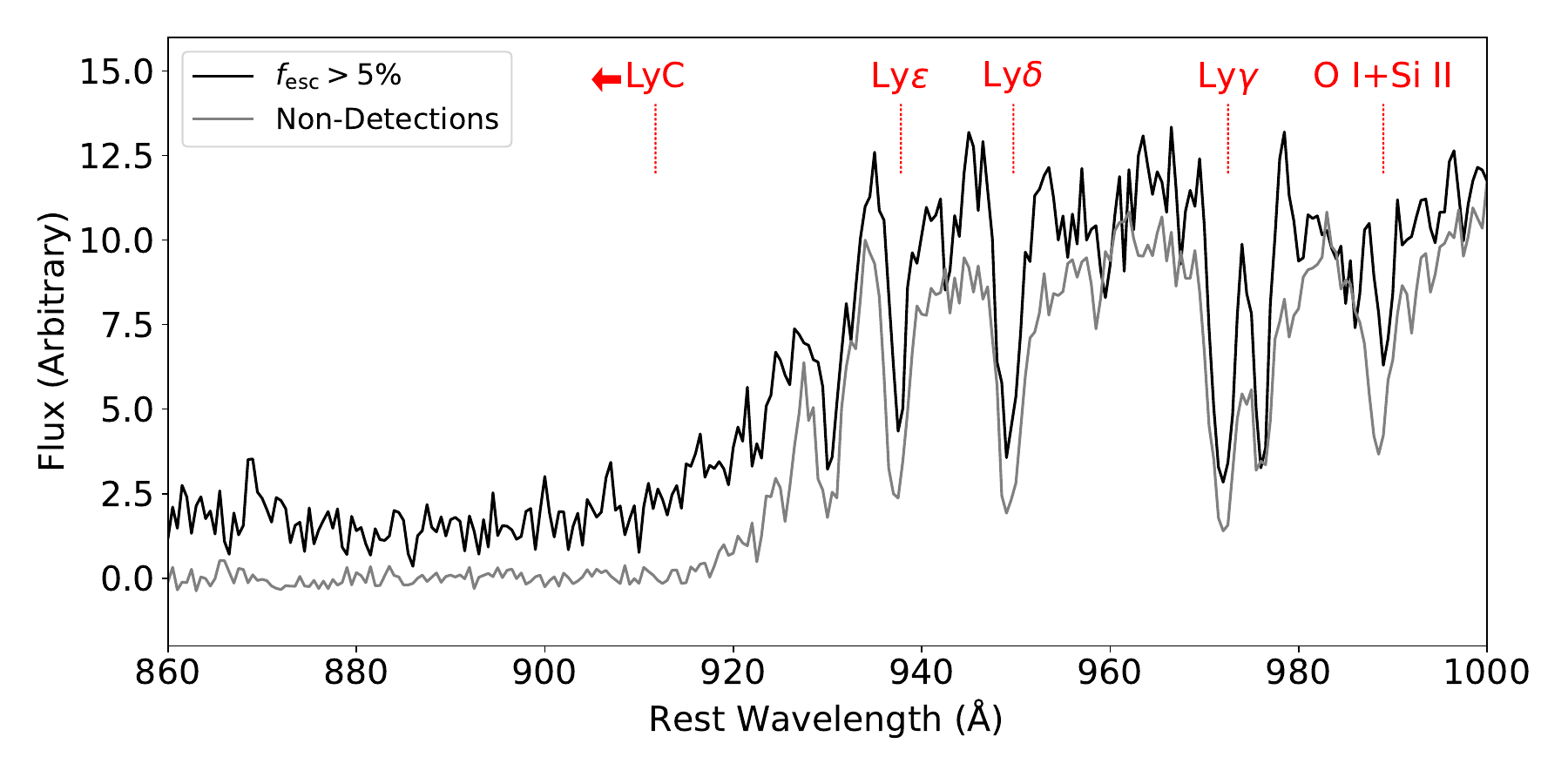}
\end{subfigure}
\begin{subfigure}[b]{0.34\linewidth}
\caption{}
\includegraphics[width=\linewidth]{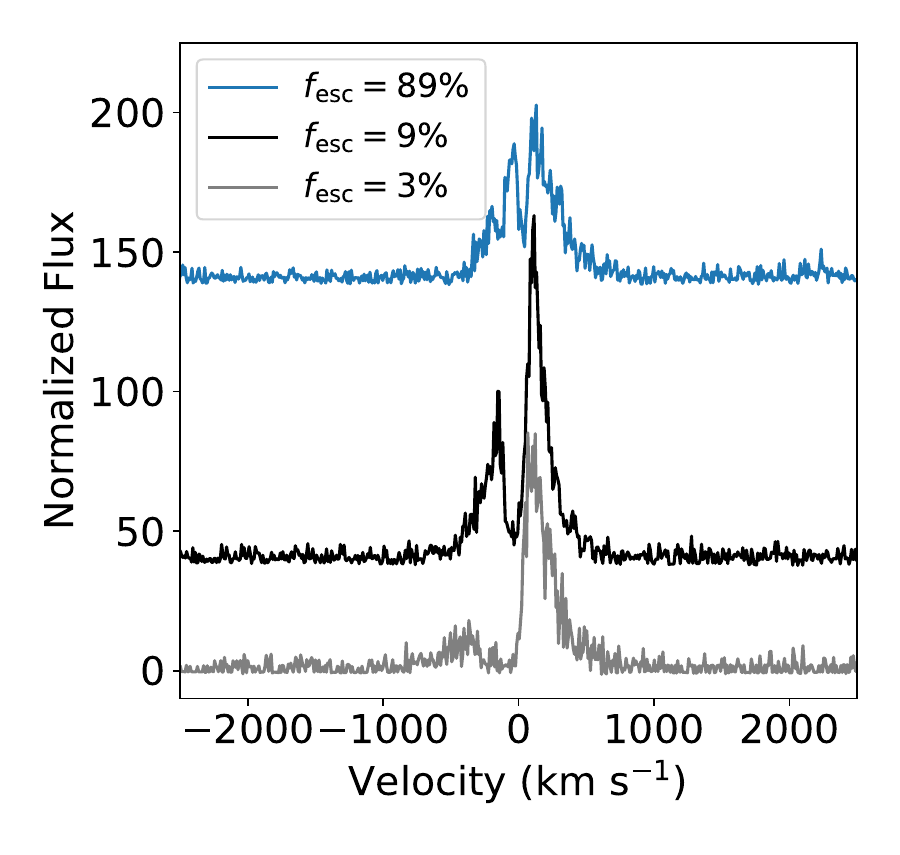}
\end{subfigure}

\caption{Panel (a) shows stacked LzLCS+ spectra of strong LCEs with \fesc\ $>5$\%\ (black) compared to non-detections (gray). The LyC region and key low-ionization absorption features are labeled in red. Strong LCEs have a clear LyC excess with a relatively flat spectral shape. Lyman series and LIS absorption lines are noticeably weaker in the strong LCEs compared to the non-detections. All stacks are normalized to the flux near 1100 \AA. Panel (b) shows example \lya\ profiles for three galaxies with different \fesc. J1243+4646 (blue) has the highest measured \fesc$=89$\%\ among the LzLCS+ and exhibits a triple-peaked profile, with narrowly separated peaks. J1011+1947 (black, \fesc$=9$\%) is typical of strong LCEs and has the median $v_{\rm sep}$ among galaxies with \fesc$>5$\%. The wider profile of J1503+3644 (gray) is typical of weak and non-LCEs and has the median $v_{\rm sep}$ for this population. The \lya\ profiles have been normalized to the continuum level and are offset for clarity. The stacks in panel (a) are described in \citet{flury24}. Data for panel (b) come from HST programs GO-14635 (PI: Y. I. Izotov) and GO 13744 (PI: T. X. Thuan) and have been described in \citet{izotov16b, izotov18b, verhamme17}.}
\label{fig:spectra}
\end{figure*}

\begin{figure*}[h]
\centering
\begin{subfigure}[b]{0.45\linewidth}
\caption{}
\includegraphics[width=\linewidth]{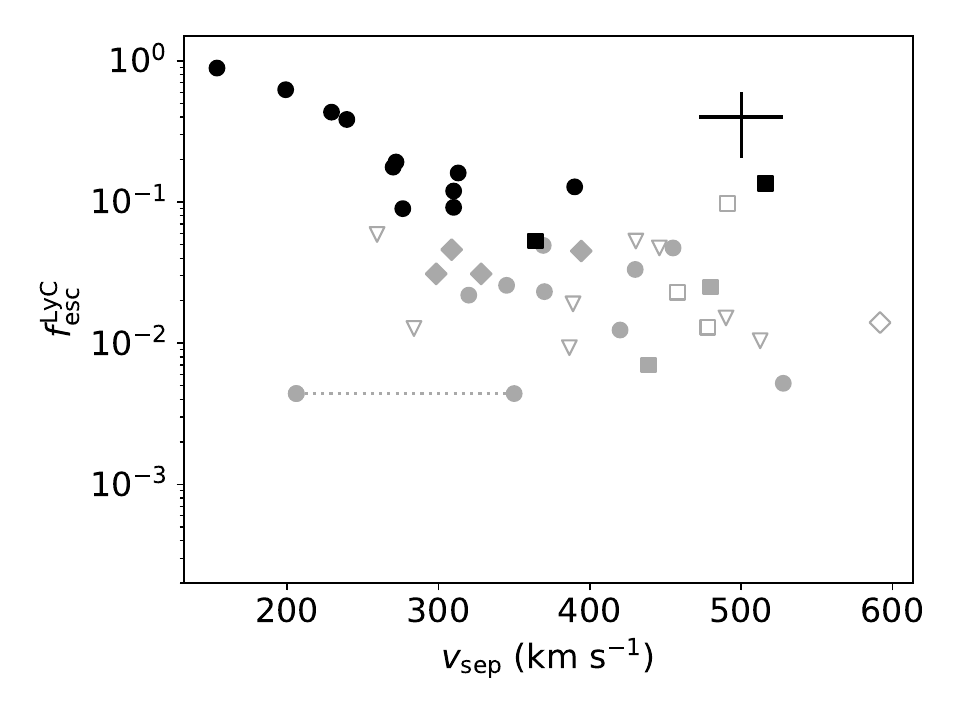}
\end{subfigure}
\begin{subfigure}[b]{0.45\linewidth}
\caption{}
\includegraphics[width=\linewidth]{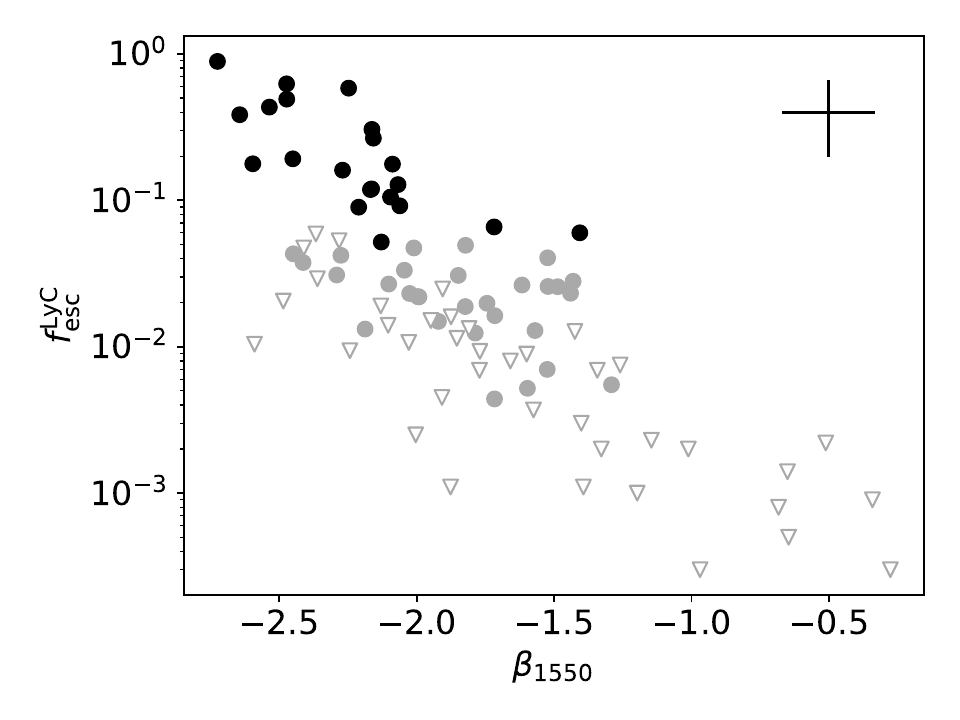}
\end{subfigure}

\begin{subfigure}[b]{0.45\linewidth}
\caption{}
\includegraphics[width=\linewidth]{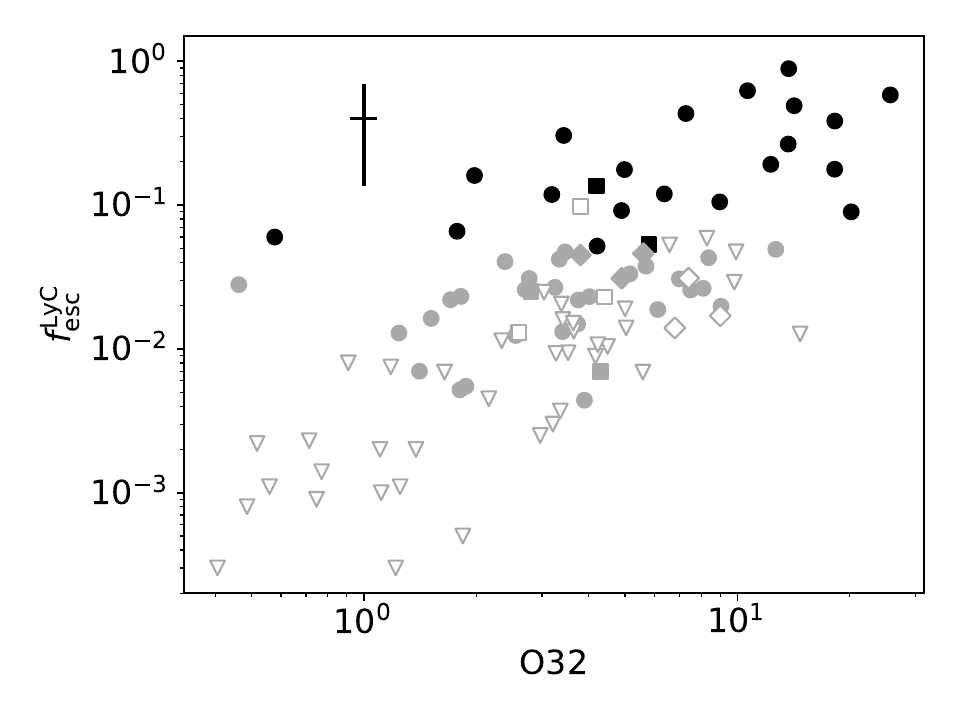}
\end{subfigure}
\begin{subfigure}[b]{0.45\linewidth}
\caption{}
\includegraphics[width=\linewidth]{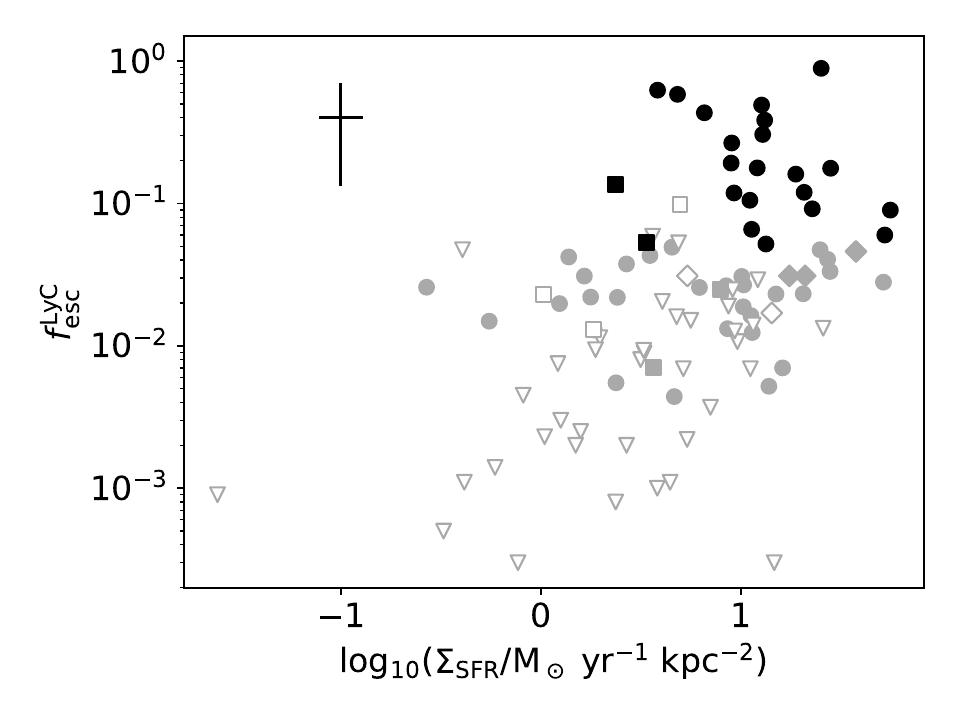}
\end{subfigure}
\caption{Four of the parameters that show the strongest correlations with \fesc\ are the separation ($v_{\rm sep}$) of the blue and red \lya\ peaks, which is sensitive to \hi\ optical depth, the UV slope $\beta_{\rm 1550}$ (b), which is sensitive to the dust attenuation, O32 (c), which traces ionization, and \sigsfr\ or related measures of galaxy compactness (d). The $v_{\rm sep}$ shows the strongest observed correlation with \fesc\ but has only been measured for a subset of galaxies with high-resolution UV spectra. The other three parameters are accessible at $z>6$ and may help predict \fesc. Black points show strong LCEs from the LzLCS+ with \fesc$>5$\%, gray points show weak LCEs with \fesc$<5$\%, and open triangles show upper limits for galaxies with non-detected LyC. Squares and diamonds indicate $z\sim0.3$ galaxies from \citet{xu22} and \citet{izotov22}, respectively, with the same shading of strong, weak, and non-detected LyC. The dashed line connects $v_{\rm sep}$ measurements for a galaxy with a triple-peaked \lya\ profile. The \fesc\ measurements from \citet{izotov22} fit the optical SED, including H$\beta$, to estimate the intrinsic LyC. The other sample \fesc\ measurements rely on UV SED fits. The H$\beta$-derived SFR estimates have not been corrected for LyC escape. Panels (a), (c), and (d) adapted from \citet{flury22b} (\copyright AAS; CC BY 4.0). Panel (b) adapted with permission from Figure 7 in \citet{chisholm22}.}
\label{fig:topfour}
\end{figure*}

\begin{figure*}[h]
\includegraphics[width=0.75\linewidth]{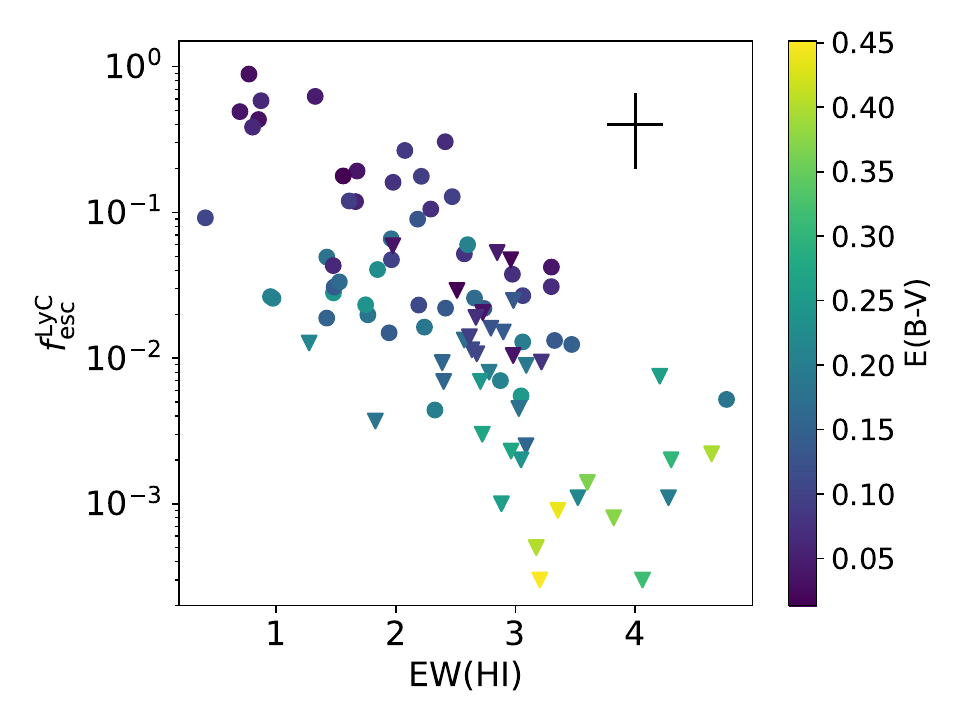}
\caption{The EW of Lyman series \hi\ absorption lines (EW(\hi)), and the UV dust attenuation E(B-V) provide the most important information for predicting \fesc\ in the LzLCS+ sample. These two variables trace the two sources of line-of-sight absorption of LyC photons: \hi\ gas and dust. At fixed EW(\hi), higher values of \fesc\ are associated with lower dust attenuation. Figure adapted from \citet{saldanalopez22} \copyright ESO, with permission.} 
\label{fig:ew_dust}
\end{figure*}

The large sample of LyC measurements at $z\sim0.3$ demonstrates that properties related to the line-of-sight ISM conditions and to stellar feedback have the strongest connection with the observed \fesc. Figure~\ref{fig:spectra} shows the UV spectral characteristics of LCEs, and Figures \ref{fig:topfour}-\ref{fig:ew_dust} show some of the strongest correlations between \fesc\ and observable quantities. The \lya\ velocity peak separation, $v_{\rm sep}$, displays the strongest correlation with \fesc\ (Figure \ref{fig:topfour}; \citealt{flury22b}) and is a proposed signature of \lya\ radiative transfer through a low optical depth medium \citep{verhamme15, dijkstra16}. However, this correlation is limited to a subset of the LzLCS+ with high-resolution spectroscopy and requires testing with a larger sample. For the full sample, \fesc\ correlates most strongly with the UV slope $\beta$ measured near 1550 \AA\ ($\beta_{\rm 1550}$; \citealt{chisholm22}), the near-UV half-light radius ($r_{\rm 50, NUV}$), and O32 (Figure \ref{fig:topfour}). Related parameters, such as other measurements of \lya\ strength (\lya\ EW and \fesclya) and compactness (\sigsfr\ and \sigsfr/$M_*$; Figure \ref{fig:topfour}), show similarly strong correlations. \citet{saldanalopez22} find that the EW and \cf\ of \hi\ and LIS absorption lines are also closely related to \fesc\ (Figure~\ref{fig:spectra}a). In a rank ordering of the variables in multivariate \fesc\ predictions, \citet{jaskot24a} find that the EW of \hi\ Lyman series absorption lines (EW(\hi)) is the single most important variable for predicting \fesc, followed by the UV dust attenuation (Figure \ref{fig:ew_dust}). Nevertheless, the correlations between \fesc\ and individual variables all show high scatter, and confirmed LCEs are diverse in their properties \citep{flury22b}. At least some of this scatter is likely due to anisotropic LyC escape; a galaxy's observed \fesc\ could be higher or lower than its true global average simply because of viewing angle.

\begin{marginnote}[]
\entry{$v_{\rm sep}$}{The velocity separation of the blue and red \lya\ emission peaks for a double-peaked profile.}
\entry{$\beta$}{The UV spectral slope, determined by assuming that flux $F \propto\ \lambda^\beta$.}
\entry{O32}{A nebular ionization ratio, defined as \oiii~$\lambda$5007/\oii~$\lambda$3727 or \oiii~$\lambda\lambda$5007,4959/\oii~$\lambda$3727 in different works.}
\entry{Star formation rate surface density (\sigsfr)}{SFR per unit area, typically defined as $\Sigma_{\rm SFR} = \frac{\rm SFR/2}{\pi r^2}$, where $r$ is the UV half-light radius.}
\end{marginnote}

\subsection{\fesc\ and Tracers of Line-of-Sight ISM Absorption}
\label{sec:lces:ism}
The observed trends with \hi\ absorption lines, $\beta_{\rm 1550}$, and \lya\ emission illustrate the natural connection between the observed LyC and the line-of-sight absorption by dust and gas. \citet{chisholm22} show that $\beta_{\rm 1550}$ predominantly traces the stellar UV dust attenuation; the close observational link between \fesc\ and $\beta_{\rm 1550}$ therefore reflects the effect of dust on \fesc. Indeed, a simple picket fence model where the \cf\ and dust attenuation regulate \fesc\ reproduces the observed \fesc, with the scatter consistent with the observational error bars \citep{saldanalopez22}. In this model, the \cf\ is inferred from the observed residual flux in the \hi\ Lyman series absorption lines. Notably, both dust absorption and \hi\ absorption are required to accurately predict \fesc\ (e.g., Figure \ref{fig:ew_dust}). Contrary to some simulations \citep[\eg][]{kimm19, mauerhofer21}, dust is apparently present within low $N_{\rm HI}$ channels and plays a role in absorbing the LyC \citep{chisholm18, saldanalopez22}. The \cf\ derived from the LIS lines correlates with the \cf\ from \hi, albeit with some scatter, and provides an alternative means of estimating \fesc\ when Lyman series lines are not available \citep{gazagnes18, saldanalopez22}. Reassuringly, the fact that the observed \fesc\ scales with the line-of-sight \hi\ and dust absorption, as expected, suggests that the calculated \fesc\ values and derived trends are reasonable, even if the exact conversion between $f_{\rm esc}$ at 900 \AA\ and the true total \fesc\ remains uncertain (Section \ref{sec:methods}). 

The resonant \lya\ line represents an alternative tracer of the \hi\ optical depth. Measurements of the \lya\ strength and line profile show some of the best correlations with \fesc, with strong LCEs (\fesc\ $> 5$\%) commonly showing high \fesclya, high \lya\ EWs, and \lya\ lines with narrow double-peaked profiles or a peak at the systemic velocity (Figure~\ref{fig:spectra}b; \citealt{verhamme17, izotov18b, flury22b, izotov22}). Nevertheless, the relationship between \fesclya\ and \fesc\ has significant scatter, with most galaxies having \fesclya\ $\geq$ \fesc\ \citep{flury22b}. Because \lya\ photons can resonantly scatter into the line of sight from other directions, they may escape even if the gas is optically thick to the LyC \citep{verhamme17}. The reverse can also be true, especially on small scales. Since the \hi\ optical depth is $10^4$ times greater in \lya\ than LyC \citep{verhamme15}, \lya\ photons can scatter even from LyC-leaking regions \citep[\eg][]{riverathorsen17a, saldanalopez23, komarova24}. For example, some strong LCEs at $z\sim0.3$ and the LyC-leaking Knot B region in Haro 11 have \fesc\ $> $\fesclya, although the two escape fractions could be marginally consistent given the uncertainties \citep{flury22b, komarova24}. 

Unfortunately, the partially neutral IGM affects \lya\ transmission at the epoch of reionization, making this indirect tracer unreliable at high redshift. \citet{henry18} and \citet{chisholm20} propose that the resonant \mgii\ doublet could serve as a proxy for the \lya\ line. The optical depth along the \mgii\ transmission paths determines the observed \mgii~$\lambda$2796/\mgii~$\lambda$2803 flux ratio \citep{chisholm20, xu22, xu23}. With this information and the calculated \mgii\ escape fraction, one can derive the \cf\ of the optically thick gas responsible for absorbing the remaining \mgii\ photons. These two constraints can then be used to estimate the \hi\ \cf\ and the residual \nhi\ in optically thin channels using information from empirical LIS-\hi\ scalings, photoionization models, and metallicity measurements \citep{xu23}. After including the additional effect of dust extinction, the estimated LyC absorption from optically thick clouds and partially transparent channels predicts the observed \fesc\ within a factor of three \citep{xu23}. The success of the \mgii\ and UV absorption line predictions \citep{xu23, saldanalopez22} demonstrate that simple models of the ISM geometry can predict \fesc\ reasonably well. As emission lines, the \mgii\ doublet may be easier to detect than the LIS absorption lines at high redshift. In addition, because the \mgii\ escape fraction tightly correlates with \fesclya\ \citep{xu23}, it can potentially constrain the \fesclya\ emerging from galaxies for studies of the high-redshift IGM \citep{henry18}.  

\subsection{\fesc\ and Global Galaxy Properties}
\label{sec:lces:global}
Although the line-of-sight ISM properties naturally show the closest connection with the observed line-of-sight \fesc, \fesc\ also correlates with more global galaxy properties. These properties give insights into the characteristics of LCEs and the mechanisms responsible for driving high \fesc. Of these properties, the strongest trends appear with nebular ionization and compactness, highlighting the importance of stellar feedback in enabling LyC escape.

\subsubsection{Ionization}
\label{sec:lces:global:ionization}
High nebular ionization is a hallmark of the strongest low-redshift LCEs. Both the observed \fesc\ and the fraction of LCEs rise with increasing O32 \citep{flury22b}, with more than 80\%\ of galaxies with O32 $>10$ exhibiting strong LyC escape (\fesc $>5$\%; \citealt{flury22b}). This high detection fraction suggests that LyC must escape at some level along most or all directions in these galaxies. However, contrary to expectations for density-bounded systems with high \fesc\ \citep[\eg][]{bergvall13, zackrisson13}, these highly ionized GPs have high EWs in H$\beta$ and other nebular lines, rather than the low EWs expected for strong isotropic escape \citep{stasinska15, schaerer18, flury22b}. The GPs' young ages and high ionizing photon production boost their intrinsic EWs, compensating for the effect of low optical depth \citep{schaerer16, schaerer18, ravindranath20, flury22b}. Furthermore, LyC escape may not be uniformly high in all directions \citep[\eg][]{jaskot24a}, such that the global \fesc\ may be lower than the most extreme observed line-of-sight \fesc\ values. 

The relationship between high ionization and \fesc\ extends to other line ratios as well. From a smaller sample of eight GPs with measured \fesc, \citet{schaerer22} show that strong LCEs have high \civ~$\lambda$1550/\ciii~$\lambda$1909 ratios. However, the existence of one GP with \fesc$\lesssim1$\%\ yet high \civ/\ciii\ shows that, like O32, high \civ/\ciii\ ratios are not a sufficient indicator of the line-of-sight \fesc. The relationship between \fesc\ and high ionization does not appear to extend to even higher ionization energies than \civ. At fixed metallicity, galaxies with higher \fesc\ tend to have higher O32 \citep{flury22b} but not higher \heii~$\lambda$4686/H$\beta$ ratios \citep{marqueschaves22a}. As a recombination line from doubly-ionized helium, requiring photons with $E>54$ eV, the lack of relationship between \heii\ and \fesc\ suggests that LyC escape does not require unusually hard ionizing sources. 
	
Whereas \oii\ emission arises from the edge of an \hii\ region, weak emission from species with even lower ionization potentials could better trace the neutral ISM and identify LCEs. \citet{wang21} find that LCEs in the LzLCS+ are systematically deficient in \sii~$\lambda\lambda$6717,6731/H$\alpha$ compared to typical SDSS star-forming galaxies. Although an \sii-based selection does result in a high fraction of LyC detections, the relationship between \fesc\ and \sii\ deficiency shows high scatter, perhaps because of anisotropic escape \citep{wang21}. Strong LCEs also appear deficient in \oi~$\lambda$6300 emission \citep{ramambason20}, but as an inherently weak line, the \oi\ diagnostic has not yet been conclusively tested on a large sample \citep{flury22b}. \hei\ emission line ratios could also trace LyC escape, due to their differing dependence on collisional excitation, fluorescence, and optical depth \citep{izotov17b}; the \hei\ diagnostic suggests that some strong LCEs may be density-bounded \citep{guseva20} but again requires confirmation on a larger sample. 

\subsubsection{Compactness}
\label{sec:lces:global:compactness}
Both \fesc\ values and the detection fraction of LCEs sharply increase for $r_{\rm 50, NUV} \leq 0.5$ kpc and \sigsfr $\geq 10$ \Msol\ yr$^{-1}$ kpc$^{-2}$ \citep{flury22b}. This threshold lies significantly above the theoretical minimum value for SN feedback to effectively clear gas via outflows \citep[\eg]{heckman01asp}. The observed connection between \fesc\ and highly concentrated star formation suggests a link between stellar feedback and \fesc. However, the specific feedback mechanism involved is unclear (see Section \ref{sec:feedback}), and the H$\beta$-based SFR used in the LzLCS+ \sigsfr\ estimate traces star formation on the shortest timescales, not the integrated effect of generations of SNe. Using a UV-based SFR, \citet{kim20} find that the related parameter \sigsfr/$M_*$ correlates with \fesclya\ and suggest that it reflects the ability of galaxies to clear channels in the ISM for both \lya\ and LyC escape. Indeed, the slightly stronger observational correlation between \fesc\ and \sigsfr/$M_*$ as opposed to \sigsfr\ \citep{flury22b} may indicate the competing role of the gravitational potential in retaining gas and preventing LyC escape.  

\subsubsection{Luminosity}
\label{sec:lces:global:luminosity}
Although the trends are weak, the LzLCS+ also shows evidence for an increase in \fesc\ toward lower $M_*$, fainter magnitudes at 1500 \AA\ ($M_{\rm 1500}$), and lower metallicity \citep{flury22b}. Taken at face value, these correlations could suggest that fainter galaxies dominate reionization. However, the scatter in \fesc\ at fixed values of these parameters is large ($\sim2$ dex), and the LzLCS+ probes a limited, relatively bright parameter space with \logten($M_*$/\Msol)=7.2-10.8 (median$=8.8$) and $M_{\rm 1500}=-18.3$ to $-21.5$ (median$=-19.9$; \citealt{flury22a}). These observed trends may largely reflect the tendency for higher O32 ratios and lower dust attenuation among the fainter galaxies in the LzLCS+ sample. In fact, multivariate analyses that take into account key variables simultaneously find that at fixed values of parameters such as O32 and $\beta_{\rm 1550}$, \fesc\ tends to increase for brighter galaxies \citep{lin24, jaskot24b}. Further LyC observations of galaxies over a wider luminosity range are necessary to discern the dependence of \fesc\ on galaxy brightness and disentangle it from other parameters. 

The weak overall trend between \fesc\ and luminosity highlights the interesting fact that LCEs occur at all luminosities and with a variety of physical properties \citep{flury22b}. Although the low-mass, low-metallicity, high O32 GPs are the strongest and most consistent LCEs, other LCE populations are more massive, metal-rich, and dusty concentrated starbursts. \citet{flury22b} suggest that to leak LyC, a galaxy must possess either high O32, high \sigsfr, or both. They further hypothesize that the cause of LyC escape may differ in these two populations, with more isotropic escape driven by radiative feedback for GPs and escape along channels carved by SN feedback in the more massive LCEs. In their ranking of variables for predicting \fesc, \citet{jaskot24a} likewise find that different variables are important for LyC escape in low- versus high-mass galaxies. In low-mass galaxies (\logten($M_*$/\Msol)$<8.8$), the EW of \hi\ absorption lines is the most important variable, with O32 providing similar information. Conversely, in high-mass galaxies (\logten($M_*$/\Msol)$>8.8$) \sigsfr\ plays the dominant role, perhaps indicating the greater importance of mechanical feedback in these systems. 

\section{INSIGHTS FROM NEARBY LCE CANDIDATES}
\label{sec:local}
At $z<0.1$, few LyC detections exist and those that do are debated (Section \ref{sec:history}). Still, these nearby galaxies can provide a detailed look at how feedback reshapes the ISM and how global LyC escape relates to smaller scale processes. In particular, mapping of galaxies at $z<0.05$ (distances $\lesssim200$ Mpc) is revealing highly ionized structures, outflows, super star clusters with diverse ages and contributions to feedback, and a complex relationship between global galaxy properties and the possible sites of LyC escape.

\begin{figure*}[h]
\centering
\begin{subfigure}[b]{0.45\linewidth}
\caption{}
\includegraphics[width=\linewidth]{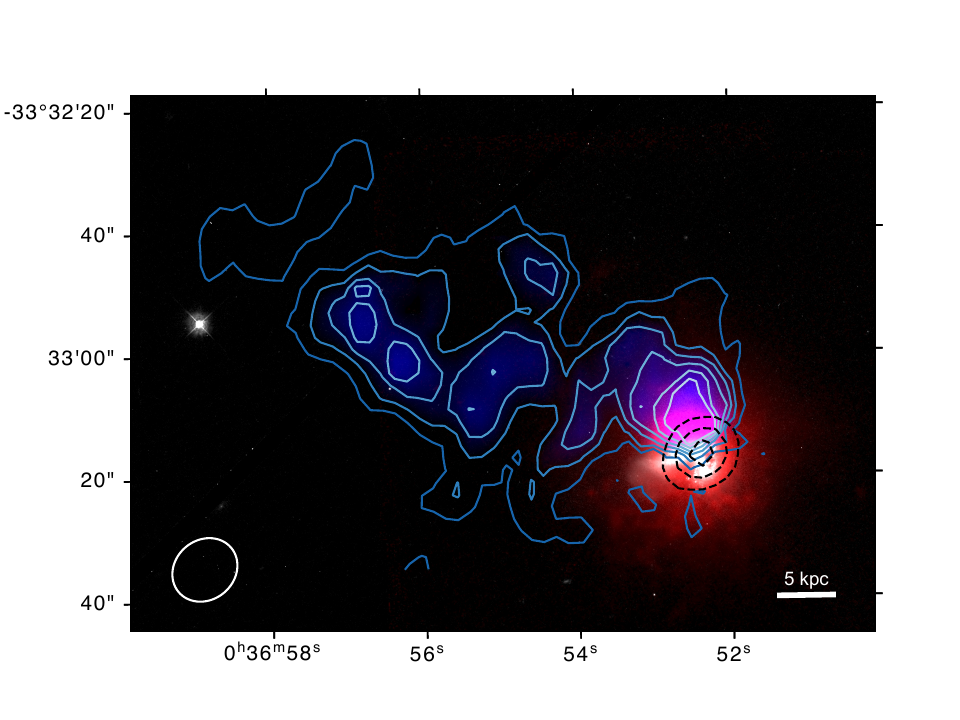}
\end{subfigure}
\begin{subfigure}[b]{0.45\linewidth}
\caption{}
\includegraphics[width=\linewidth]{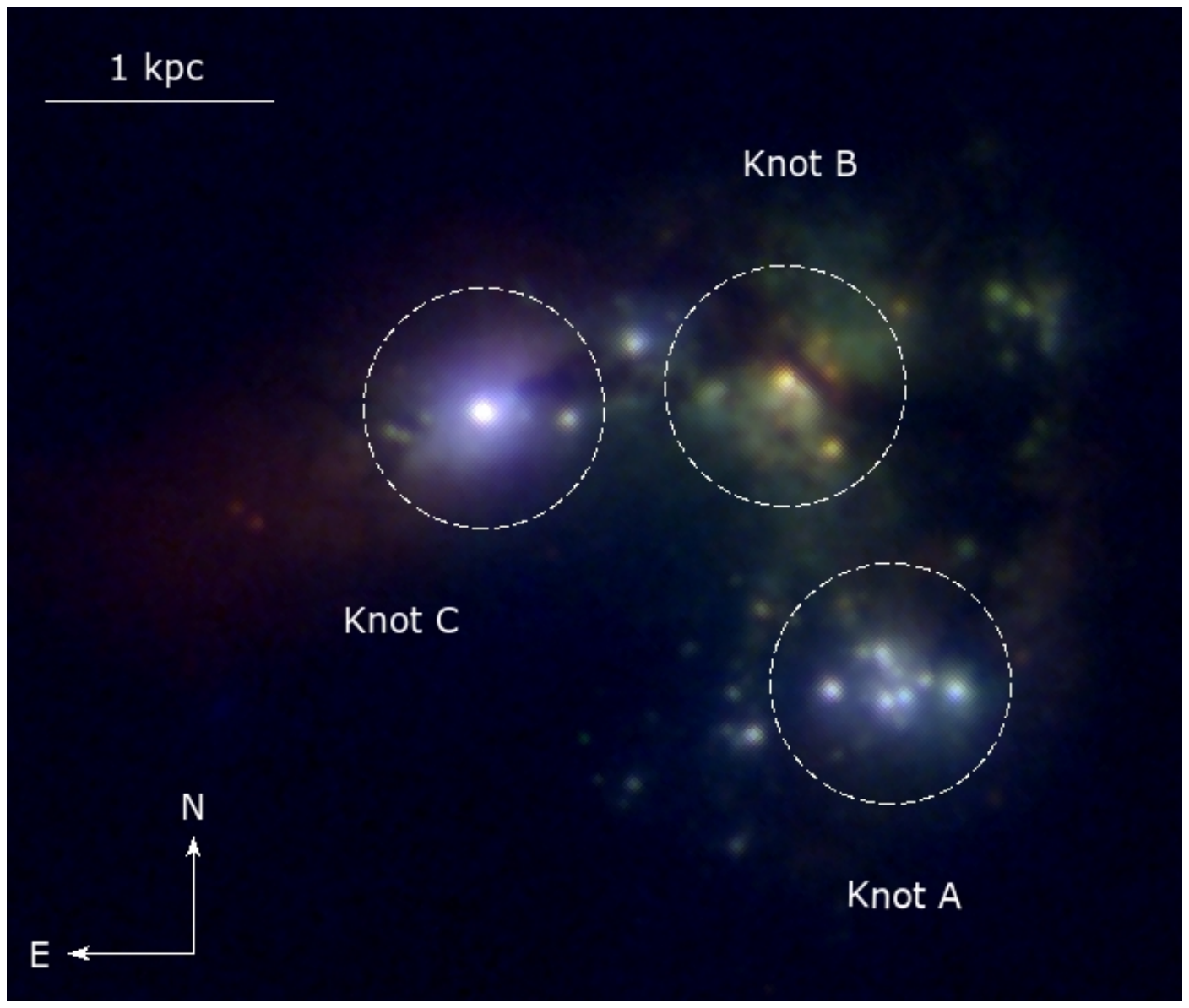}
\end{subfigure}

\begin{subfigure}[b]{0.3\linewidth}
\caption{}
\includegraphics[width=\linewidth]{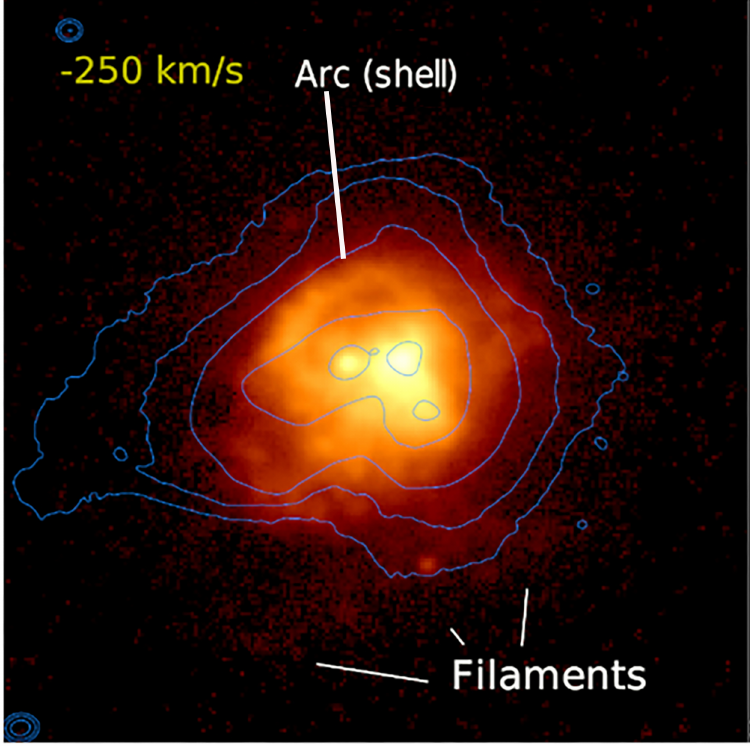}
\end{subfigure}
\begin{subfigure}[b]{0.3\linewidth}
\caption{}
\includegraphics[width=\linewidth]{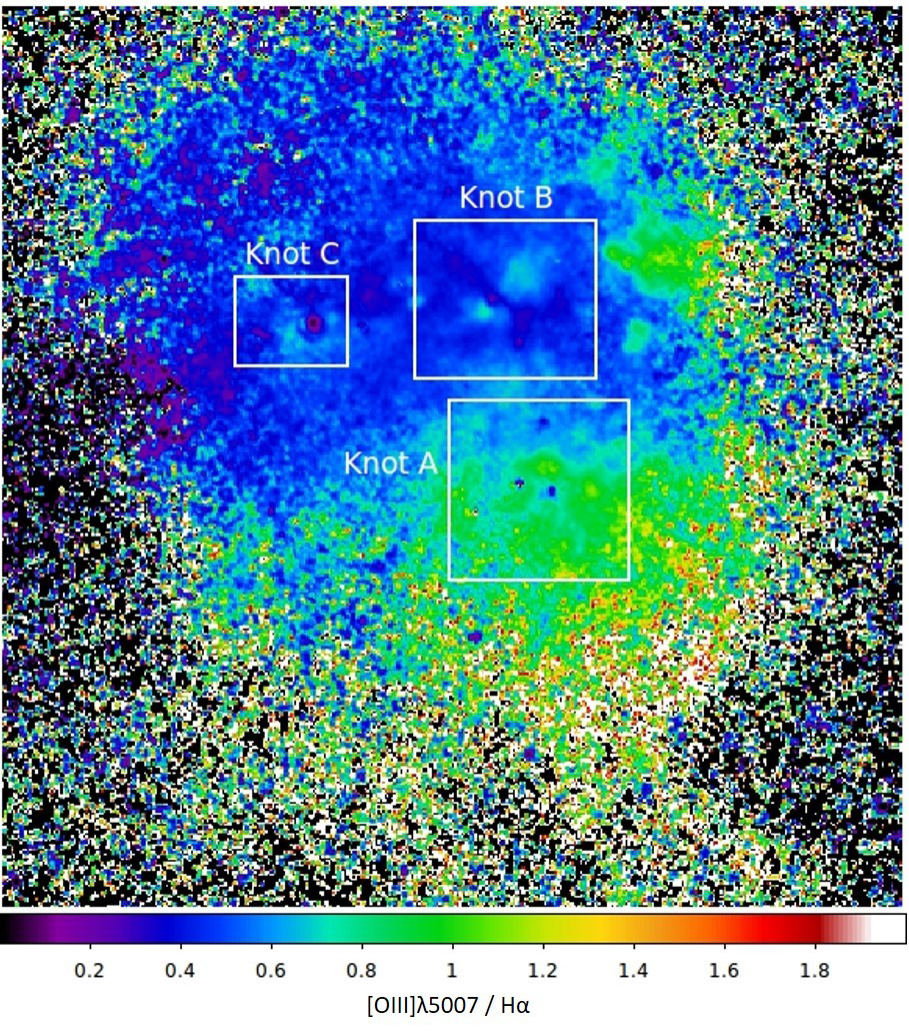}
\end{subfigure}
\begin{subfigure}[b]{0.3\linewidth}
\caption{}
\includegraphics[width=\linewidth]{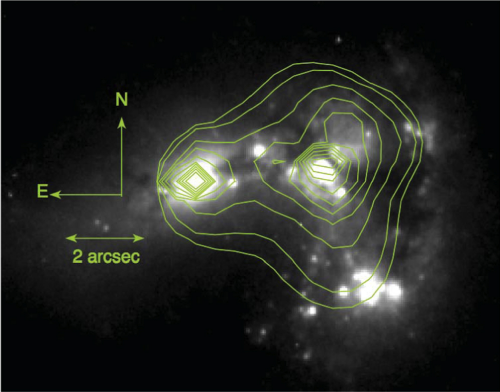}
\end{subfigure}
\caption{Haro 11, a nearby LBA with detected LyC, exhibits complex ISM structures and contradictory indirect LyC signatures. Panel (a) shows extended, displaced \hi\ 21 cm emission (blue with solid contours) resulting from a merger. The image also shows the synthesized beam (white ellipse), H$\alpha$ (red), optical stellar emission (white), and 21 cm absorption (dashed black contours). Panel (b) identifies Haro 11's three star-forming knots, with {\it HST} images of the 5500 \AA\ continuum, 3360 \AA\ continuum, and \lya\ shown in red, green, and blue. The \lya\ predominantly comes from Knot C. In Panel (c), a Multi-Unit Spectroscopic Explorer (MUSE) map of H$\alpha$ gas moving at -250 km s$^{-1}$ reveals a disjointed superbubble shell near Knot C and filaments extending from Knot A. Contours show {\it I}-band continuum. In Panel (d), {\it HST} \oiii/H$\alpha$ maps show ionized channels near Knot B and a large ionized region around Knot A. {\it Chandra} 0.3-5 keV X-ray contours appear atop an {\it HST} optical image in Panel (e). X-ray point sources coincide with Knots B and C, and diffuse X-ray emission extends throughout the star-forming region. Panel (a) adapted with permission from Figure 1 in \citet{lereste24}. Panel (b) reproduced from \citet{komarova24} (\copyright AAS; CC BY 4.0). Panel (c) adapted with permission from Figure 3 in \citet{menacho19}. Panel (d) reproduced from \citet{keenan17} by permission of the AAS. Panel (e) reproduced from \citet{prestwich15} by permission of the AAS. 
}
\label{fig:haro11}
\end{figure*}

The LBA Haro 11 is the most well established nearby LCE (but see Section \ref{sec:history}), yet the origin of its LyC escape has long been a mystery. Haro 11 contains three star-forming knots, denoted A, B, and C, each of which is a candidate LCE based on different criteria (Figure \ref{fig:haro11}). Knots A and B possess significant young populations ($\leq 4$ Myr; \citealt{komarova24}). Their high nebular O32 and low line-of-sight \oii/H$\alpha$ ratios suggest possible LyC escape, with Knot A also showing high ionization and possible LyC escape in transverse directions \citep{keenan17}. On the other hand, the low \fesclya$\lesssim1$\%\ and high \cf\ from LIS lines (0.95-0.96 at line center) imply high optical depths \citep{ostlin21}. Of the three knots, Knot A displays the broadest \lya\ emission. Knot B has a narrower \lya\ line profile but also shows underlying \lya\ absorption from higher \nhi\ $\sim10^{21}$ cm$^{-2}$ gas and higher dust reddening \citep{ostlin21}. The strongest outflow originates from Knot A \citep{sirressi22}, but as the oldest star-forming region, Knot C has experienced the strongest SN feedback resulting in a high gas velocity dispersion \citep{menacho21}. It has the highest \fesclya\ and lowest LIS \cf\ (0.55), with a narrow \lya\ profile \citep{ostlin21}. However, it has lower O32 \citep{keenan17} and high reddening \citep{ostlin21}. Both Knots B and C contain X-ray sources \citep{prestwich15}, which may contribute additional feedback. Figure \ref{fig:haro11} highlights key properties of Haro 11 from multi-wavelength imaging. 

Recent COS observations of all three knots detected LyC from Knots B and C, with \fesc$= 0.034 \pm 0.029$ and $0.051 \pm 0.043$, respectively \citep{komarova24}. These results support \lya\ peak separation as an indicator of \fesc\ but demonstrate that even regions with low \fesclya\ may have escaping LyC. The observations also highlight the importance of LyC luminosity \citep{komarova24}. By virtue of its young age and high LyC photon production, Knot B dominates the total LyC flux and luminosity, despite its lower \fesc. Given the high observational uncertainties, these observations cannot rule out LyC escape from Knot A (the 2$\sigma$ limit is \fesc$<0.1$) or definitively rank the three knots by \fesc. Regardless, Haro 11 illustrates that LyC escape in LCEs may be complicated on small scales, with multiple regions contributing to \fesc. The regions that contribute the most LyC flux, such as Knot B, may not be same regions that dominate the global measurement of proposed indirect diagnostics. For instance, Knot C dominates the observed \lya\ flux, whereas Knots A and B generate most of the H$\alpha$ and \oiii\ emission \citep{ostlin21}. 

Observations of the ISM in Haro 11 and other nearby starbursts clarify the physical processes that may contribute to LyC escape.  \hi\ 21 cm maps show that a galaxy merger has displaced the \hi\ gas 6 kpc away from the star-forming regions \citep{lereste24}. By both triggering star formation and redistributing absorbing neutral material, mergers may aid LyC escape \citep{lereste24}. Feedback may further alter the ISM geometry. \citet{menacho19} identify highly ionized channels emanating from Knot B, a fragmented supershell centered around Knot C (radius $\sim1.7$ kpc), and an even larger fragmented superbubble (radius $\sim3.3$ kpc) from older star formation episodes $\sim15$ Myr ago that may promote LyC escape near the younger Knot A. 

In other galaxies, similar ISM structures highlight the anisotropic nature of LyC escape and the possible role of successive generations of star formation in boosting \fesc. Using IPM, \citet{zastrow11, zastrow13} detect narrow ionization cones with opening angles of $40^{\circ}$ in two nearby starburst dwarf galaxies. The ionization cones originate from concentrated star-forming regions, but non-detections in other starbursts show that high \sigsfr\ alone does not guaranteed the formation of ionized channels. Rather, they may also require an optimal age range or star formation history, with SNe first generating outflows and sufficient remaining or newly formed stars supplying the ionizing photons \citep{zastrow13}. The latter, multi-stage starburst scenario appears to characterize several other galaxies with hints of LyC escape. In the possible LCE Tol 1247-232, feedback from a 12 Myr-old burst has dispersed interstellar gas, producing superbubbles and filaments and potentially helping a younger, 2-4 Myr-old generation of stars  to ionize large regions of the ISM \citep{micheva18}. Like Haro 11, a galaxy merger may have also contributed to triggering star formation and disturbing the ISM \citep{micheva18}. Similarly, the LCE candidate ESO 338-IG04 exhibits bipolar ionization cones associated with feedback-driven outflows; LyC from the youngest clusters likely escapes anisotropically along these channels \citep{bik18}. 

Galaxies like Haro 11 can illuminate the ISM conditions in galaxies with moderate \fesc$<5$\%, but they may not perfectly capture the physical properties of more extreme galaxies like the GPs. Haro 11 and Tol 1247-232 are both more massive (\logten($M_*$/\Msol)$=9.7-10.1$; \citealt{leitet13, chisholm15}) and less ionized (O32$ = 1.5-3.4$; \citealt{verhamme17}) than most LCE GPs (median \logten($M_*$/\Msol)=8.5, O32$\geq5$; \citealt{flury22a}). Similarly, the burst in ESO 338-IG04 is less extreme (H$\alpha$ EW$= 570$ \AA; \citealt{ostlin09}) than the GPs (H$\alpha$ EW$\sim1000$ \AA; \citealt{izotov16a, izotov16b, izotov18a, izotov18b}. The nearby starburst NGC 2366/Mrk 71 is a well-studied template of the extreme ionization conditions in the GPs albeit in a $\sim100$ times less luminous galaxy \citep{micheva17}. 

\begin{figure*}[h]
\centering
\begin{subfigure}[b]{0.7\linewidth}
\caption{}
\includegraphics[width=\linewidth]{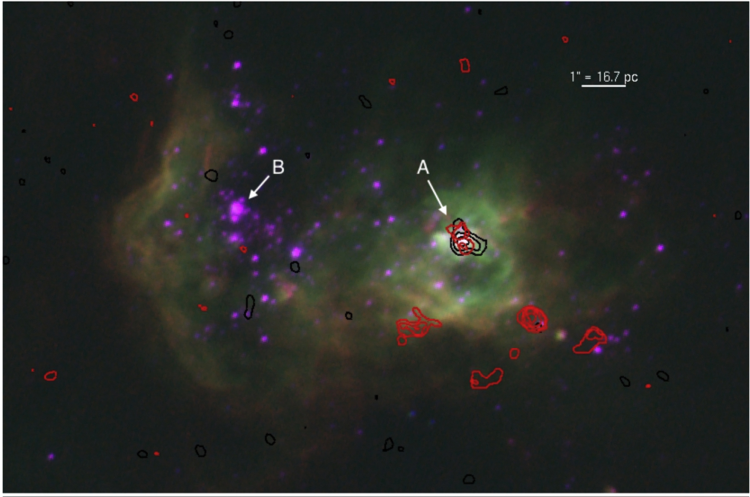}
\end{subfigure}

\begin{subfigure}[b]{0.25\linewidth}
\caption{}
\includegraphics[width=\linewidth]{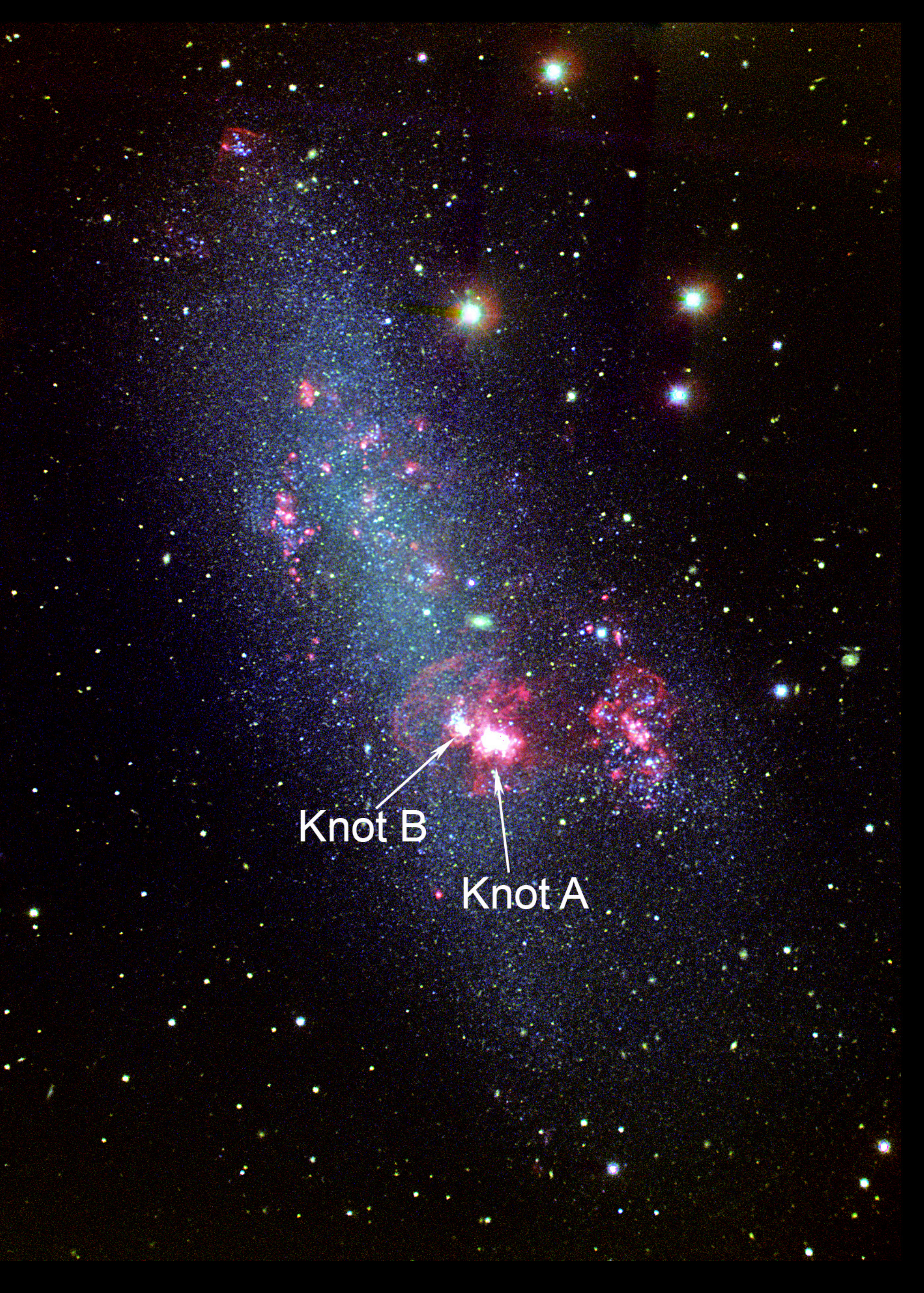}
\end{subfigure}
\begin{subfigure}[b]{0.5\linewidth}
\caption{}
\includegraphics[width=\linewidth]{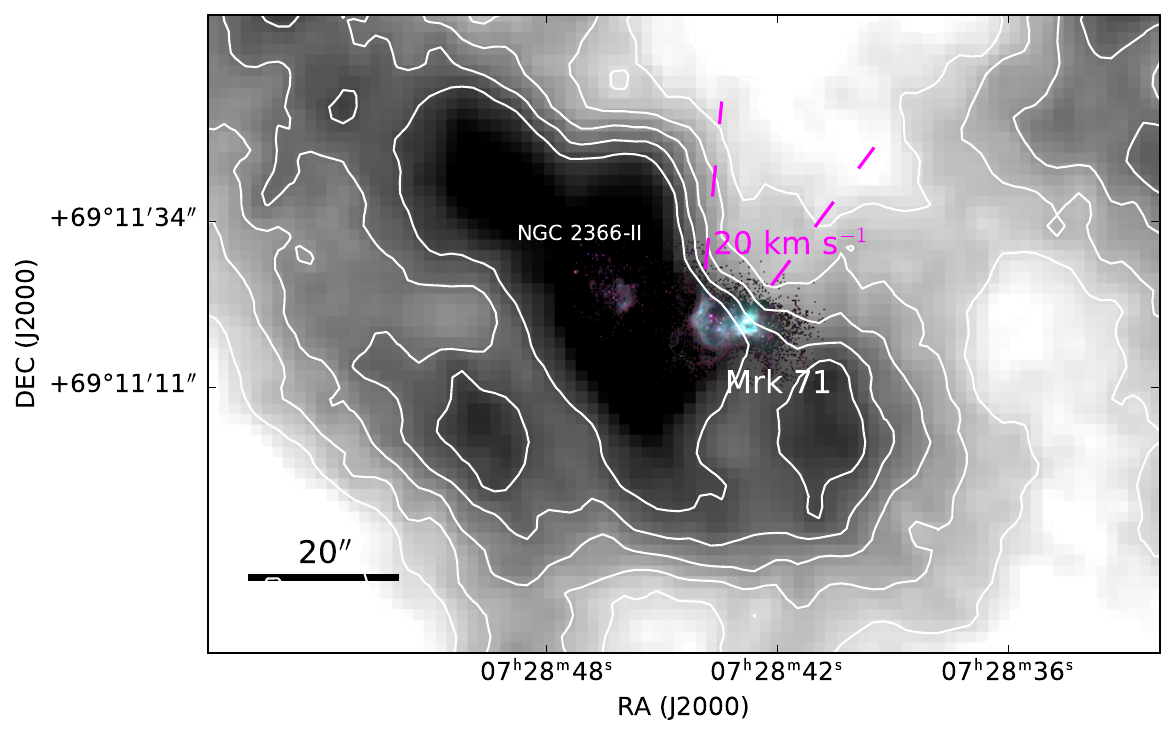}
\end{subfigure}

\caption{Panel (a) shows an {\it HST} image of the GP-like star-forming region, Mrk 71, with \oii~$\lambda$3727 emission in red, \oiii~$\lambda$5007 in green, and blue stellar continuum light in blue. The younger (1 Myr old) Knot A is extremely ionized and compact, while the older (3-5 Myr), neighboring Knot B shows a superbubble morphology. Red contours show CO(2-1) emission near the young Knot A, and black contours show radio continuum emission. The arrow identifying Knot B also indicates the approximate position of a {\it Chandra} X-ray point source \citep{kaaret24}. As shown in Panel (b), Mrk 71 is a small bright region within the fainter, more extended galaxy NGC 2366. {\it B}-band, {\it R}-band, and H$\alpha$ emission are shown in blue, green, and red, respectively. In Panel (c), a grayscale \nhi\ map is overlaid on Mrk 71's optical nebular emission. A 20 km s$^{-1}$ blowout region of low \nhi\ extends from Mrk 71. Panel (a) reproduced from \citet{oey17} by permission of the AAS. Panel (b) image credit Observatorio de Calar Alto, J. van Eymeren (AIRUB, ATNF) and A. L{\'o}pez S{\'a}nchez (Macquarie University) with permission. Panel (c) reproduced from \citet{micheva17} by permission of the AAS. }
\label{fig:mrk71}
\end{figure*}

The Mrk 71 starburst region in the low-mass galaxy NGC 2366 matches the GPs in O32 and metallicity and would appear as a compact GP if located at a greater distance (Figure \ref{fig:mrk71}; \citealt{micheva17}). Mrk 71 consists of two super star clusters, Knot A and Knot B. The extremely young ($\sim1$ Myr old) Knot A shows evidence for Very Massive Stars \citep{smith23}, produces 90\%\ of the ionizing luminosity, and is responsible for Mrk 71's GP-like properties \citep{micheva17}. Consistent with its young age, Knot A retains significant neutral gas, showing both nearby molecular gas \citep{oey17} and deep \lya\ absorption, although neutral material may not fully cover the ionizing source \citep{smith23}. Radiative feedback may dominate Knot A, with catastrophic cooling in dense gas limiting the effectiveness of mechanical feedback \citep{oey17}. In contrast, the older Knot B (3-5 Myr) has formed a superbubble with a blowout and outflow \citep{micheva17}, possibly driven by an X-ray source \citep{kaaret24}. NGC 2366 is thus another candidate for a two-stage starburst scenario, with mechanical feedback from Knot B aiding the escape of LyC photons produced by the younger Knot A \citep{micheva17}. 

Multi-wavelength studies of $z<0.3$ GPs have revealed the unique characteristics of these prototypical LCEs. At the lowest redshifts ($z\lesssim0.05$), O32-selected galaxies are sometimes called ``blueberry galaxies" for their blue appearance in SDSS images; this high O32 population extends to lower stellar masses and metallicities than the $z\sim$0.3 LCEs \citep{yang17c}. UV and optical spectra of GPs with the highest O32 $>20$ confirm that GP properties are linked to extremely young starburst ages ($\lesssim3$ Myr; \citealt{izotov17b, jaskot17}) with high ionization parameters and high ionizing photon production rates \citep{izotov17a}. The diversity of \lya\ and LIS absorption line properties observed in the GPs indicate that LyC escape is likely direction dependent; while some GPs have narrow \lya\ emission and weak LIS lines, others show deep absorption features or even narrow \lya\ emission within an underlying absorption trough \citep{henry15, yang17b, mckinney19, jaskot19, izotov20}. Like Mrk 71 Knot A, these young starbursts may still retain significant neutral material (see Sections \ref{sec:geometry} and \ref{sec:feedback}). Also like Mrk 71, young GP starbursts may be intensely bright regions, lying within an older, possibly more extended galaxy \citep[\eg][]{izotov11, paswan22b}. \hi\ 21 cm observations show that some GPs, especially those with the highest O32, may be globally deficient in \hi\ for their SFR \citep{mckinney19, kanekar21, dutta24}, suggesting that high star formation efficiencies could also be a potential factor in LyC escape. Major mergers may disturb their \hi\ gas and trigger these extreme starbursts \citep{purkayastha22, dutta24}. Finally, X-ray emitting sources could contribute to feedback in the GPs, with possible ultraluminous X-ray sources or low-luminosity AGN detected in some, but not all, GPs \citep[\eg][]{svoboda19, harish23, borkar24, adamcova24, singha24}. 

Nearby galaxies also present an opportunity to probe LyC escape in faint galaxies. \citet{choi20} investigate \fesc\ in NGC 4214, which has a similar mass and metallicity as the GPs but contains a fainter, less extreme starburst ($M_{\rm FUV}\sim-16$). By comparing the predicted ionizing photon production from resolved stellar populations with the observed emission from ionized gas and dust, \citet{choi20} conclude that \fesc\ varies considerably (0-40\%) in different locations and estimate a global \fesc\ $=25^{+16}_{-15}$\%. Whereas young stellar populations characterize the GPs \citep[\eg][]{izotov17b, izotov18b}, in NGC 4214, the highest \fesc\ $=20-40$\% and highest intrinsic ionizing luminosities are associated with superbubble structures. The youngest stellar populations exhibit lower \fesc\ $=0-24$\%. NGC 4214 shows that LyC may escape even from less extreme star-forming galaxies, and the sites of LyC escape may differ from those in the GPs.

These studies of nearby star-forming galaxies and GPs hint at the possible factors that may contribute to LyC escape. Many galaxies show evidence of multi-stage starbursts, suggesting that mechanical feedback works in concert with radiation from the youngest burst to generate high \fesc. \hi\ observations point to the role of mergers in disrupting the ISM and imply high SFR to \hi\ mass ratios in candidate LCEs. X-ray sources also appear present in many potential LCEs. While suggestive, further research is necessary to establish the connection between these factors and \fesc, especially as these nearby galaxies may not be the exact counterparts of the confirmed LCEs at $z=0.3$. 

\section{ISM GEOMETRY IN LCES}
\label{sec:geometry}
Simple models of a density-bounded or picket fence gas geometry approximate but imperfectly describe real LCEs. In theory, O32 could trace density-bounded \hii\ regions, yet the strengths of low-ionization emission lines in GPs and LCEs are incompatible with pure density-bounded photoionization models \citep[\eg][]{stasinska15, plat19, ramambason20}. Some optically thick clumps or sight-lines must exist. Likewise, even highly ionized GPs show Lyman series and LIS absorption lines \citep[\eg][]{gazagnes20, mckinney19, izotov20}, implying at least partial covering fractions of neutral gas. The high observed H$\beta$ EWs ($\sim150-300$ \AA) in the GPs also demonstrate that \fesc\ cannot be uniform in all directions, as the high amount of required LyC absorption would seemingly contradict the highest observed values of \fesc\ ($>50$\%; e.g., \citealt{zackrisson13, flury22b}). Assuming lower global \fesc\ values with excursions to extreme \fesc\ along particular sight-lines would alleviate this discrepancy. 

\begin{figure*}[h]
\includegraphics[width=0.9\linewidth]{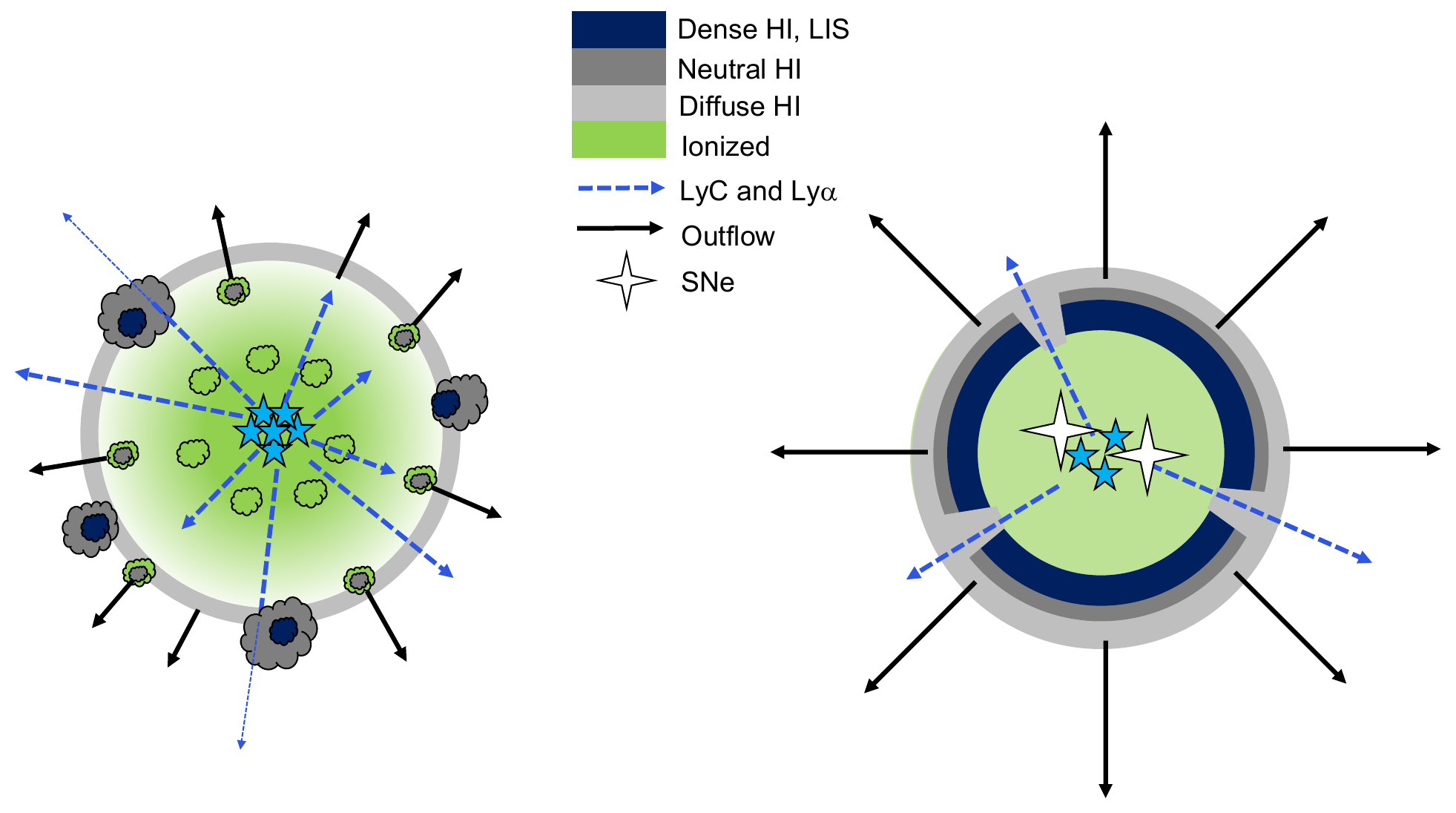}
\caption{Left: In strong LCEs, LyC escape is nearly isotropic, although \nhi\ variations lead to varying \fesc\ in different directions. At young ages, gas may be turbulent and clumpy. LIS absorption lines trace the \cf\ of the densest gas (dark blue), whereas \hi\ absorption lines have a higher \cf\ and trace both optically thick and optically thin \hi\ (gray). Radiation is the dominant feedback mechanism. High ionization causes low neutral column densities in many directions, while radiation incident on partially neutral clumps and low \nhi\ channels accelerates them to high velocities.  Right: Weaker LCEs may be slightly older bursts dominated by mechanical feedback. With a weaker ionizing radiation field, more gas is neutral, and stellar winds and SNe have pushed material into a superbubble morphology. Holes in the superbubble shell transmit LyC photons. The LIS and \hi\ \cf\ are more similar to each other, and fast outflows with high mass loading are detectable in the neutral gas. These geometric pictures combine insights and models from e.g., \citet{heckman01, jaskot19, gazagnes20, komarova21, carr24, flury24, menon24}. }
\label{fig:geometry}
\end{figure*}

A picket fence model including a uniform dust screen reproduces the observed \fesc\ for most LCEs (Section \ref{sec:lces}), yet systematically under-predicts the \fesc\ of the strongest leakers \citep{gazagnes20, saldanalopez22}. \citet{gazagnes20} propose that absorbing clumps in the strongest LCEs may be partially optically thin, representing a hybrid of the pure picket-fence and pure density-bounded scenarios. Although optically thick to Lyman series emission, these clouds would still transmit some fraction of the incident LyC. The differing inferred \cf\ of the \hi\ versus the LIS absorption lines supports this geometric picture. At both low and high redshift, the LIS line \cf\ ($C_{\rm f, LIS}$) appears systematically lower than than the \cf\ inferred from \hi\ lines ($C_{\rm f, HI}$; \citealt{reddy16b, saldanalopez22}). Using stacked spectra from the LzLCS+, \citet{flury24} explain that this \cf\ discrepancy is unrelated to metallicity and instead reflects a changing gas geometry. The $C_{\rm f, LIS}$ represents the \cf\ of the densest, most optically thick clouds. Because of the high opacity to Lyman series photons, $C_{\rm f, HI}$ traces absorption from both higher and lower column density material, including clouds that are optically thin to the LyC. At higher \fesc, both the \cf\ of dense clouds and the average \nhi\ between these clouds declines; in the LzLCS+ stacks, this effect manifests observationally as a decreasing $C_{\rm f, LIS}/C_{\rm f, HI}$ along with a decreasing EW(\hi) from the lower \nhi\ \citep{flury24}. Figure \ref{fig:geometry} illustrates the possible geometry of the strongest LCEs. 

The above constraints imply that \fesc\ is anisotropic, due to a patchy gas distribution. Yet, in the strongest LCEs, LyC may still escape in most directions, albeit with varying \fesc\ depending on the $C_{\rm f, HI}$ and \nhi\ along a particular sight line \citep[\eg][]{gazagnes20}. The high fraction of LyC detections among the GPs \citep{izotov16b, izotov18b, flury22b} indicates that this population likely does have escape in most directions, as the GPs were not explicitly selected for a favorable orientation. In a multivariate analysis of \fesc\ predictors, \citet{jaskot24a} find that global properties do not distinguish the highest \fesc\ GPs from more weakly leaking GPs. Instead, the most important property for correctly predicting the highest \fesc\ values is the EW(\hi),  tracing line-of-sight \hi\ absorption. The dominant cause of \fesc\ variations among the GPs may therefore be random orientation. Maps of nebular emission in the strongest LCEs also imply LyC escape in most directions. Strong LCEs appear equally compact in \oii\ and the lower ionization \mgii\ line, implying the absence of extended neutral circumgalactic gas \citep{leclercq24}. Consistent with the picture of global escape but line-of-sight \fesc\ variation, weak global \oi\ and \sii\ emission distinguishes strong LCEs but characterizes some lower \fesc\ galaxies as well \citep{ramambason20}. In summary, the observed line-of-sight \fesc\ appears to be set by a combination of the dense gas \cf, the residual \nhi\ between clumps, and dust absorption, with the strongest LCEs deficient in dense neutral gas along most sight lines. Stellar feedback may play an essential role in reducing \cf\ and \nhi\ and setting up conditions favorable for LyC escape. 

\section{FEEDBACK MECHANISMS AND LYC ESCAPE}
\label{sec:feedback}
Many simulations and theoretical arguments suggest that SN-driven outflows are critical for LyC escape, as they may disperse natal neutral gas and carve channels in the ISM \citep[\eg][]{heckman01, clarke02, heckman11, trebitsch17, barrow20}. Yet observational support for this mode of LyC escape is less clear. From UV spectra of luminous LBAs, \citet{heckman11} and \citet{alexandroff15} find that candidate LCEs and one confirmed LCE have faster outflows and more concentrated star formation, as expected from SN-driven LyC escape. Furthermore, as discussed in Section \ref{sec:local}, resolved studies of local LCEs and candidates reveal fragmented superbubbles and ionized conical outflows, which may lead to LyC escape along particular sight-lines. 

\subsection{Feedback in Strong LCEs: The Role of Radiation}
\label{sec:feedback:radiation}
The GPs, the strongest known LCE population, present a more complex picture, however. High-resolution spectra of emission lines in a subset of the LzLCS+ sample suggest that strong LCEs do have higher-velocity, preferentially blue-shifted ionized gas emission \citep{amorin24}. Nevertheless, galaxies span an order of magnitude in \fesc\ at fixed line width. Absorption line observations suggest that fast outflows along the line of sight do not guarantee high \fesc\ and vice versa. Nearby, highly ionized GPs show signatures of LyC escape, such as weak LIS absorption and strong, narrow \lya, yet have low outflow velocities, even $\sim0$ km s$^{-1}$ in one case \citep{jaskot17}. Outflows traced by absorption lines are present in confirmed LCE GPs, but their velocities are no more extreme than those of a control sample \citep{chisholm17}. Modeling of the \mgii\ doublet in the LzLCS+ sample likewise suggests that strong, neutral outflows are not implicated in LyC escape \citep{carr24}. Most strong LCEs lack detectable outflows in \mgii. When outflows are present, strong LCEs show lower velocities and weaker mass, energy, and momentum loading factors compared to weak and non-LCEs. \citet{carr24} argue that outflows with high mass loading of neutral gas would impede LyC escape, as the neutral outflow would naturally result in high \cf. The fast outflows with high mass loading inferred for the weaker LCEs might correspond to the classical scenario of SN-driven outflows propelling neutral clouds and enabling LyC escape along narrow channels \citep{carr24}. 

Outflow driving mechanisms may vary across the LCE population, with the dominant type of feedback dependent on the age of the stellar populations. Some GPs possess emission lines with sloping, power-law shaped wings that may originate from radiative feedback \citep{komarova21}. Indeed, the high ionization conditions in GPs appear associated with extremely young ages, prior to SN onset \citep[\eg][]{izotov17b, izotov18b, smith23}. Radio spectra of strong LCEs show a flat spectral index, likewise consistent with thermal emission from young \hii\ regions and dense ionized gas, rather than SNe \citep{bait24}. Moreover, at the low metallicities of GPs ($12+$\logten(O/H)$\lesssim8.0$), weaker stellar winds and a higher fraction of direct-collapse black holes may delay SN feedback to ages $>10$ Myr and establish radiative feedback as the dominant mechanism \citep{jecmen23}. 

Weak mechanical feedback, combined with the compact, dense conditions in GPs, may lead to catastrophic cooling, preventing the formation of superbubbles and instead generating a clumpy gas geometry that permits LyC escape \citep{jaskot19, jecmen23, carr24}. Young ages, intense radiation, and dense clouds may explain the high O32 ratios and ionization parameters of the GPs and connection of these properties with high \fesc\ \citep{jaskot19}. Both the GP-like Mrk 71 Knot A and the LCE Sunburst cluster at $z=2.4$ show evidence of dense, pressurized gas clouds near a young ionizing cluster \citep{smith23, pascale23}. While some dense material may remain near the cluster, lighter, partially-neutral clumps can be accelerated to high velocities by absorbing LyC and \lya\ photons traveling through low \nhi\ channels; these radiatively-driven outflows could account for the broad, power-law wings observed in GP emission lines (Figure \ref{fig:geometry}; \citealt{komarova21}). \citet{amorin24} propose that broad emission line wings could also originate from turbulence, which may be present even at young ages. 

In addition to driving outflows, the GPs' intense radiation may enhance LyC escape by directly ionizing large portions of the ISM. Strong ionizing radiation may destroy dust grains, one source of LyC opacity \citep{chisholm18}. This effect alone cannot explain high \fesc, however, as high O32 correlates with increased \fesc\ even at fixed dust attenuation \citep{jaskot24a}. In this multivariate analysis, O32 emerges as an important predictive parameter when EW(\hi) and \lya\ emission variables are excluded, which suggests that ionization and neutral gas properties are related and somewhat interchangeable. A clumpy ISM structure and young ages may work together to ensure LyC escape. While ISM measures like EW(\hi) correlate with \fesc, the fraction of light from the youngest ($<3$ Myr) population is the strongest factor in predicting high \fesc\ in a multivariate analysis of stacked LzLCS+ spectra \citep{flury24}. Strong radiation may reduce the \nhi\ between dense clouds \citep[\eg][]{jaskot19}, and stacked spectra with higher inferred $\xi_{\rm ion}$ tend to show lower EW(\hi) \citep{flury24}. Photoionization modeling of nearby dwarf starbursts also suggests that high ionizing photon production per unit stellar mass is linked to the highest \fesc\ but possibly over a wider age range (2-6 Myr; \citealt{ramambason22}.)

Nevertheless, radiation may not be solely responsible for strong \fesc. Stacked LzLCS+ spectra reveal that older, $8-10$ Myr, populations contribute significant emission in strong LCEs, and the expected mechanical feedback from these populations correlates with reduced \cf\ \citep{flury24}. Hence, a two-stage burst may optimize LyC escape, with SNe forming an inhomogeneous ISM (Section \ref{sec:local}) while radiation from young stars further reduces the column densities of clumps and channels via ionization. Notably, while strong mechanical feedback seems to separate LCEs from non-LCEs, only the presence of substantial young populations distinguishes strong LCEs from weaker LCEs \citep{flury24}. 

\subsection{Feedback in Weak LCEs: Mechanical Feedback and Supernovae}
\label{sec:feedback:mechanical}
SNe may instead dominate the feedback in weaker LCEs. Stacked LzLCS+ spectra suggest that intermediate age ($3-6$ Myr) populations are prevalent in galaxies with \fesc$=1-5$\%\ \citep{flury24}. Weak leakers also show steeper radio spectra, suggestive of cosmic ray acceleration by SNe \citep{bait24}. By sweeping up neutral clumps into a superbubble shell, mechanical feedback from stellar winds and SNe may reduce \fesc\ by creating a more homogeneous ISM structure \citep[\eg][]{fujita03, jaskot19}. Cosmic rays produced in SN remnants can similarly smooth the ISM \citep{farcy22}, and declining ionizing photon production at older ages may increase the neutral gas fraction. Holes within superbubbles may still provide paths for weak levels of \fesc\ \citep[\eg][]{fujita03, menacho19}. Consistent with this picture, UV spectra of LAEs suggest that ISM structure and kinematics vary with age. The youngest LAEs exhibit low \cf, low outflow velocities, higher O32, and more \lya\ escape, with evolution to higher \cf, higher velocities, and weaker escape at older ages \citep{hayes23, hayesetal23}. Figure \ref{fig:geometry} summarizes the possible differences in gas geometry and feedback in strong versus weak LCEs. 

\subsection{Feedback from X-ray Sources}
\label{sec:feedback:xray}
Accreting black holes may be another influential source of both mechanical and radiative feedback in LCEs. Although low metallicities may delay SN feedback \citep{jecmen23}, X-ray binaries from direct collapse black holes can provide mechanical feedback once the most massive stars end their lives. Ultra-luminous X-ray Sources are present in nearby LCE candidates such as Haro 11, Tol 1247-232, Tol 0440-381, and ESO 338-IG04 \citep{prestwich15, kaaret17, kaaret22, oskinova19} and often coincide with the confirmed or suspected locations of LyC escape \citep[\eg][]{micheva18, komarova24}. On the other hand, harder ionizing sources are not preferentially associated with confirmed or candidate LCEs \citep{marqueschaves22b, ramambason22}. 

In the GPs, X-ray emission may be over-luminous relative to the SFR-metallicity-luminosity relation, follow the expected relation, or be absent altogether, depending on the object. Over-luminous X-ray emission has been detected in three GPs, although two of these are less ionized and more massive than most confirmed GP LCEs \citep{svoboda19, adamcova24}. Some GPs have candidate low-luminosity AGN \citep{harish23, borkar24, adamcova24}, but most low-mass GPs actually appear to be under-luminous in X-rays, perhaps due to their young burst ages \citep{adamcova24}. The X-ray source in Mrk 71 Knot B \citep{kaaret24} demonstrates that X-ray sources may not always belong to the dominant ionizing stellar cluster in GPs but could still contribute relevant mechanical feedback. Evidence for the role of X-ray sources in LyC escape is currently inconclusive but warrants investigation with larger samples of GPs and confirmed LCEs. The detection of X-ray emission in several candidate LCEs demonstrates that X-ray feedback could be a relevant piece of the feedback budget in these galaxies. 

\section{PREDICTING LYC ESCAPE AT HIGH REDSHIFT}
\label{sec:highz}
The \fesc\ trends observed at low redshift offer a means of predicting \fesc\ at $z>6$, where the IGM interferes with direct LyC detection. Based on $z\sim0.3$ observations, likely LCEs are $z>6$ galaxies with blue UV slopes, highly ionized nebular gas, and/or compact star formation. Given the diversity of LCEs (Section \ref{sec:lces}), galaxies with less extreme properties could also contribute to ionizing the IGM at a weaker level. Several works have proposed diagnostics based on the variables that show the strongest correlations with \fesc, including the \lya\  profile shape \citep{izotov18b, naidu22} and $\beta_{\rm 1550}$ \citep{chisholm22}. High observed O32 has also been used to claim likely LyC escape \citep[\eg][]{williams23}. While promising, these correlations provide an incomplete picture. With the possible exception of the limited \lya\ $v_{\rm sep}$ observations, the relationships between \fesc\ and a single variable typically show high scatter, with \fesc\ spanning 1-2 dex at a fixed parameter value \citep{flury22b}. Furthermore, as discussed in Section \ref{sec:geometry}, LyC escape may depend on multiple aspects of the ISM geometry, including the \cf, \nhi\ between dense clumps, and the dust absorption, whereas a single variable is unlikely to capture this complexity. For instance, the scatter in the \fesc-$\beta_{\rm 1550}$ relation may reflect changing \hi\ optical depth at fixed dust attenuation (Figure \ref{fig:ew_dust}; \citealt{chisholm22}). Conversely, properties like \lya\ profile shape and O32 may trace \hi\ conditions more than dust absorption \citep[\eg][]{verhamme15, hayesetal23, jaskot24a}. 

The use of indirect diagnostics assumes that the relationship between galaxy properties and \fesc\ does not evolve with redshift. This assumption has yet to be fully tested, but strong LCEs at $z\sim2-4$ do share certain properties with their low-redshift counterparts. Like the GPs, the strong LCE Ion 2 has a high O32 $=7$, negligible dust attenuation, and compact effective radius $=340$ pc \citep{vanzella16, vanzella20}. The LyC-leaking region within the lensed, $z=2.37$ Sunburst Arc also shares several distinctive properties with the GPs. Blueshifted emission lines indicative of outflows demonstrate feedback at work \citep{mainali22}, similar to strong low-redshift LCEs \citep{amorin24}. Compared to the rest of the galaxy, the LyC-leaking region has a younger age, bluer $\beta$ slope, higher O32 and \oiii\ EW, greater \fesclya, and concentrated star formation \citep{kim23}, which suggests that these properties are connected to \fesc\ at both low and high redshift. The starburst age is $\sim 3-4$ Myr \citep[\eg][]{mainali22, riverathorsen24}, with H$\alpha$ emission implying strong radiation from a $<4$ Myr old population \citep{riverathorsen24}. Both the GP LCEs and the LCE Sunburst Arc cluster also possess enhanced N/O \citep{izotov23, pascale23, riverathorsen24}. Like the strongest GP LCE, the \lya\ emission from the Sunburst Arc, Ion 2, and other $z=2-4$ LCEs shows a peak at or near the line center, consistent with escape through a low \nhi\ hole \citep{riverathorsen17b, vanzella20, naidu22}. Stacks of LAEs with narrow \lya\ profiles also share remarkably similar properties with GPs, such as high O32, high nebular EWs, strong \civ\ emission, high \fesclya, and optically thin \mgii\ emission \citep{naidu22}. LyC measurements in stacked spectra of $z=3-4$ star-forming galaxies agree that \fesc\ is enhanced in galaxies with smaller radii \citep{marchi18}, higher \lya\ EWs \citep{steidel18, marchi18}, and narrower \lya\ profiles \citep{pahl24}, although these relationships may differ quantitatively from those measured at low redshift. 

The observed \fesc\ for $z\sim2-4$ galaxies does not always agree perfectly with low-redshift predictions, especially for predictions relying on a single diagnostic variable. LyC escape in stacked LBG spectra appears shifted to higher \fesc\ at fixed \lya\ EW and fixed \lya\ red peak velocity compared to the LzLCS+ \citep{saldanalopez23, pahl24}. Due to low resolution, these spectra cannot test the much tighter predicted relationship with \lya\ $v_{\rm sep}$. It is not clear if these discrepancies reflect evolution in the diagnostics or a difference in some other galaxy properties between the low- and high-redshift samples. For instance, the $z\sim3$ LBGs are both more luminous and redder than most of the strong LCEs in the LzLCS+, which may instead be better analogs of high-redshift emission-line galaxies. These differences suggest a need for multivariate \fesc\ diagnostics, as other galaxy properties may affect the observed relationships between \fesc\ and individual variables. Similarly, LyC non-detections in $z\sim2-4$ stacked spectra  \citep{rutkowski17, naidu18} have lower O32, lower nebular EWs, and higher stellar masses than strong low-redshift LCEs \citep{izotov18b}. \citet{citro24} likewise report non-detections for LyC in seven lensed $z\sim2$ galaxies, despite their blue $\beta$ ($\sim-2.4$ to -2) and moderately high \lya\ EWs (10-100 \AA). These galaxies' \fesc\ limits are considerably lower than most LzLCS+ galaxies with similar $\beta$. On the other hand, their \fesc\ limits are consistent with low-redshift samples given their \lya\ EWs \citep{chisholm22, flury22b}. Other high-redshift galaxies with redder $\beta$ slopes imply higher \fesc\ at fixed $\beta$ than predicted from the limited low-redshift observations in this regime \citep{saldanalopez23}. The existing high-redshift constraints demonstrate the need for caution in using $\beta$ alone to infer \fesc, especially since this parameter traces only one component of the LyC absorption.

\begin{figure*}[h]
\centering
\begin{subfigure}[b]{0.45\linewidth}
\caption{}
\includegraphics[width=\linewidth]{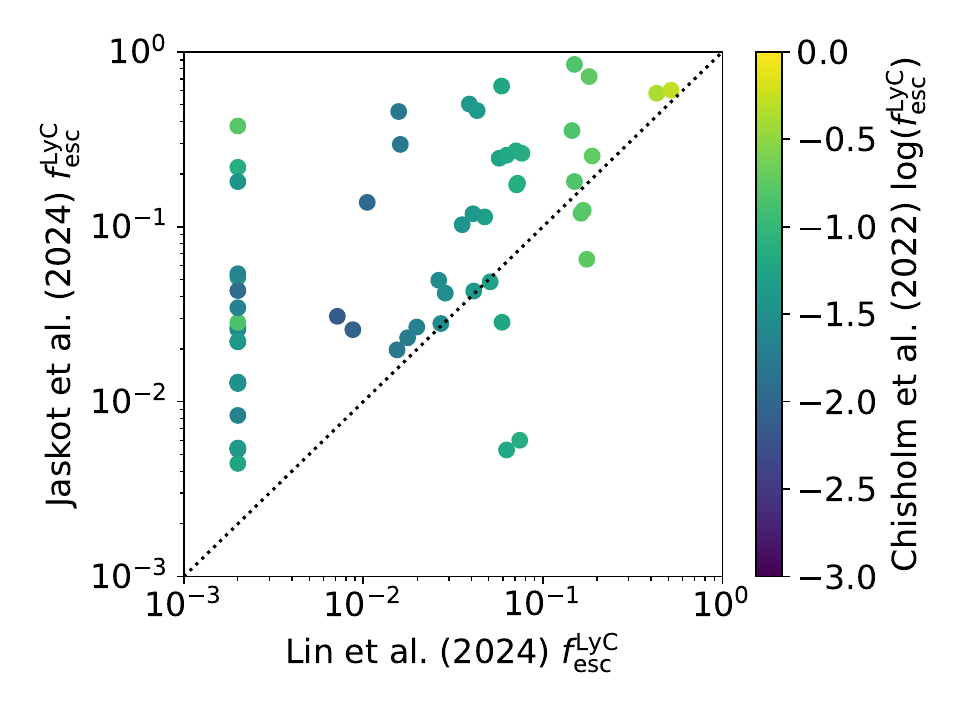}
\end{subfigure}
\begin{subfigure}[b]{0.45\linewidth}
\caption{}
\includegraphics[width=\linewidth]{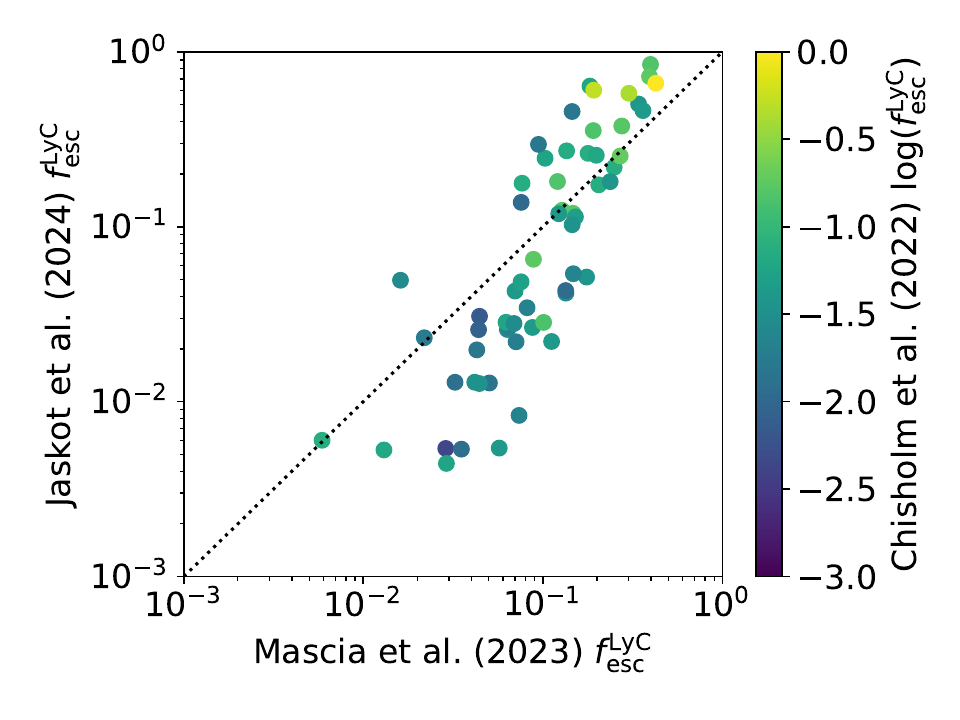}
\end{subfigure}

\begin{subfigure}[b]{0.45\linewidth}
\caption{}
\includegraphics[width=\linewidth]{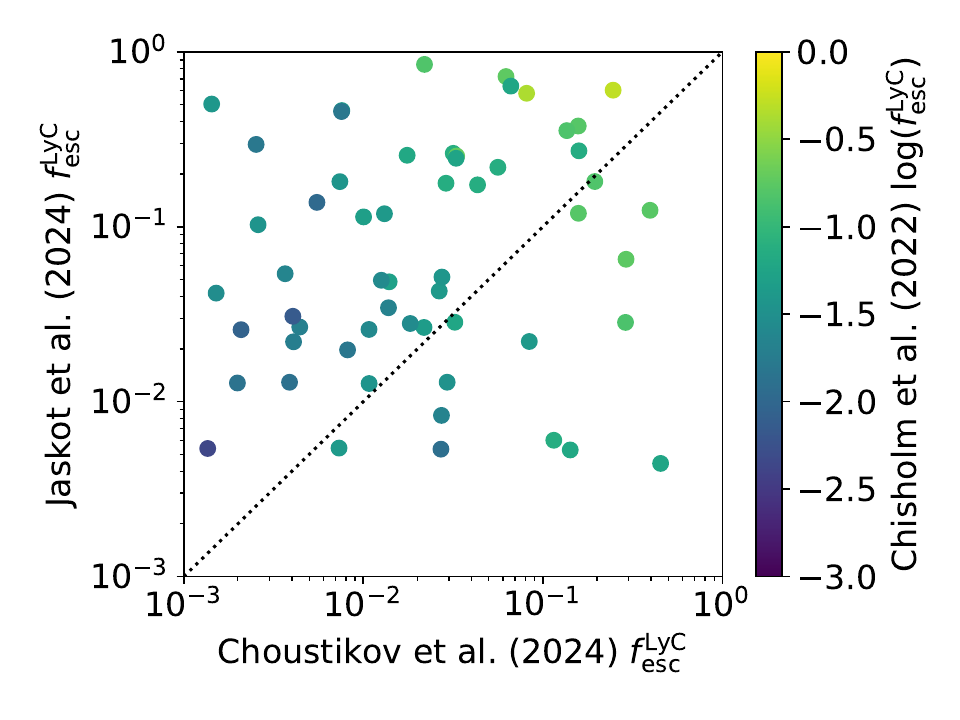}
\end{subfigure}
\begin{subfigure}[b]{0.45\linewidth}
\caption{}
\includegraphics[width=\linewidth]{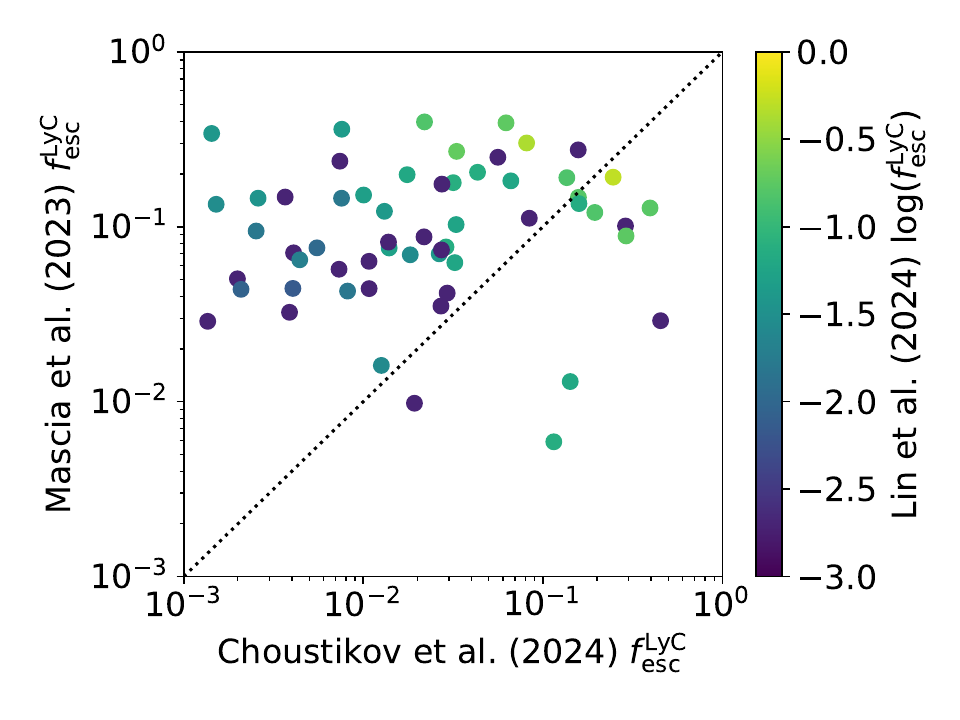}
\end{subfigure}
\caption{A comparison of \fesc\ predictions for $z>4$ galaxies from \citet{mascia23, mascia24}. Indirect diagnostics based on different methods and sets of variables \citep{chisholm22, mascia23, lin24, jaskot24b, choustikov24a} can give conflicting predictions. The dashed line shows one-to-one agreement. For the probabilistic \citet{lin24} method, we determine \fesc\ for galaxies with a $\geq50$\%\ probability of being an LCE, and plot non-leakers (\fesc\ $=0$) at \fesc\ $=0.002$. }
\label{fig:highz}
\end{figure*}

Given the high scatter in single variable diagnostics and the dependence of \fesc\ on multiple parameters, several works have proposed diagnostics based on combinations of multiple variables. Diagnostics combining dust attenuation with LIS absorption or \mgii\ emission lines attempt to constrain LyC absorption from both the neutral gas and dust along the line of sight (Section \ref{sec:lces:ism}; \citealt{gazagnes18, chisholm18, chisholm20, saldanalopez22, xu22, xu23}). \citet{mascia23} use the LzLCS+ to derive a relation between three of the strongest predictive variables (O32, UV radius, and $\beta$) and \fesc. Because this method does not treat non-detections as upper limits, it may overestimate \fesc\ for weak LCEs. To account for non-detections, \citet{lin24} propose using $M_{\rm 1500}$, $\beta$, and O32 to first predict the probability of a galaxy being an LCE, afterwards using the \fesc-$\beta$ relation \citep{chisholm22} to derive \fesc\ for the LCEs. \citet{jaskot24a, jaskot24b} have created a flexible \fesc-prediction tool, customizable to any set of measurements available for high-redshift and LzLCS+ galaxies. Importantly, this multivariate model uses a survival analysis method, which incorporates information from both LyC detections and non-detections. Since all of these multivariate methods \citep{mascia23, lin24, jaskot24a} derive from the LzLCS+ sample, they agree on the general properties of high \fesc\ galaxies. However, their different treatment of upper limits and inclusion or omission of \sigsfr, radius, and $M_{\rm 1500}$ can lead to significant differences in their \fesc\ predictions, especially for moderate or weak LCEs (Figure \ref{fig:highz}; \citealt{jaskot24b}). Multivariate predictions can also deviate substantially from predictions based on a single variable (Figure \ref{fig:highz}; \citealt{mascia24, jaskot24b}.)

Some of these multivariate diagnostics have been tested on galaxies at intermediate redshift. \citet{saldanalopez23} use $C_{\rm f, LIS}$ plus UV dust attenuation to predict \fesc\ for $z=3-5$ galaxies from the VANDELS survey. The leaky galaxies identified by these ISM tracers resemble low-redshift LCEs in many ways, with higher \fesc\ found for galaxies with lower masses, bluer $\beta$, stronger \lya\ emission, higher $\xi_{\rm ion}$, and high \civ/\ciii. These predictions correctly identify two confirmed LCEs and reproduce the observed \fesc\ of one of the two. \citet{jaskot24b} also apply their multivariate models to $z\sim3$ LyC samples, with some success but some caveats. Most models work equally well for high and low-redshift galaxies, which suggests that low-redshift \fesc\ diagnostics could indeed hold at high redshift. However, the most predictive variables (e.g., O32 and \sigsfr) have only been tested on small numbers of $z\sim3$ galaxies, and the applicability of low-redshift diagnostics is not yet definitive. Models incorporating O32, dust attenuation, UV luminosity, and morphology information appear the most promising but require further testing. 

With the recent deluge of $z>6$ JWST observations, many studies have begun using indirect diagnostics to predict \fesc\ for galaxies in the epoch of reionization. These \fesc\ predictions differ in detail but all find a plethora of candidate LCEs. Using $\beta$ slopes, \citet{atek24} estimate \fesc$ = 4.5-16$\%\ for ultrafaint, lensed galaxies with $M_{\rm UV}=-15$ to -17. With these galaxies' high observed $\xi_{\rm ion}$, their moderate \fesc\ values imply that such faint galaxies could dominate reionization. By applying their multivariate model to JWST samples at $z=4.5-9$, \citet{mascia23, mascia24} find an average \fesc\ of 12-13\%, with a minority of galaxies (20\%) having \fesc\ $> 20$\%.  Using their survival analysis models, \citet{jaskot24b} find a similar \fesc\ distribution, with most galaxies showing low \fesc\ (median 5-14\%) but with a subset of the population extending to higher \fesc\ values. However, the \citet{jaskot24b} \fesc\ estimates have high uncertainty, and the more reliable models were only applied to smaller samples with the necessary spectroscopic data. Neither \citet{jaskot24b} nor \citet{mascia24} find a strong trend of \fesc\ with $M_{\rm UV}$, but these preliminary trends need confirmation with larger spectroscopic samples as well as testing the effect of different input variables on the \fesc\ predictions. \citet{lin24} apply their multivariate model by assuming a distribution of $\beta$ and O32 for $z=8$ galaxies. Their results suggest that \fesc\ peaks at intermediate luminosities ($M_{\rm UV}=-16$ to -19) but depend on the adopted O32 and $\beta$ trends. For all of the above predictions, the modest \fesc\ for faint galaxies imply that they could dominate reionization, thanks to their higher $\xi_{\rm ion}$ and greater numbers \citep{atek24, lin24, mascia24, jaskot24b}. To verify these \fesc\ estimates and associated conclusions, we must also understand whether or how the various indirect \fesc\ diagnostics evolve with redshift and with galaxy luminosity.

While intermediate-redshift results tentatively suggest that some low-redshift models can predict \fesc, recent JWST observations provide further evidence of similarities between $z\sim0.3$ LCEs and reionization-era galaxies. Observed galaxies at $z=4.5-8$ share similar emission line properties, $\beta$, UV luminosities, and compactness as low-redshift metal-poor starbursts and follow the same emission line trends with stellar mass \citep{schaerer22b, mascia23}. Nevetheless, the use of indirect diagnostics still faces several challenges. The LzLCS+ sample covers $M_{\rm UV}=-18.5$ to -21.5, and \fesc\ predictions for faint ($M_{\rm UV} > -18$) galaxies \citep[\eg][]{atek24, lin24} therefore rely on extrapolation. Certain extreme galaxies at $z\sim3$ and $z > 6$ do not have analogs among the low-redshift samples \citep[\eg][]{marqueschaves22b, bunker23}, and many $z>6$ galaxies appear more compact than the LzLCS+ \citep{jaskot24b}. The applicability of \fesc\ diagnostics in these regimes is unknown. Despite these caveats, results for galaxies sharing the low-redshift parameter space suggest \fesc$\geq10$\%\  may be common at high redshift \citep[\eg][]{mascia24}. Combined with evidence for high $\xi_{\rm ion}$ and unexpected numbers of $z>9$ galaxies, the existing \fesc\ estimates imply that the observed galaxy population may actually provide too many ionizing photons \citep{munoz24}. The true average \fesc\ and $\xi_{\rm ion}$ at $z>6$ may need to be much lower than current estimates. An anticorrelation between \fesc\ and $\xi_{\rm ion}$ could also resolve the problem \citep{munoz24}, but low- and high-redshift observations suggest the opposite may be true \citep[\eg][]{schaerer16, guseva20, ramambason22, saldanalopez23}. Alternatively, a higher IGM clumping factor could alleviate the tension \citep{davies24}. The puzzling LyC excess at $z>6$ highlights the need to better test the validity of low-redshift correlations in the high-redshift regime. 

\section{COMPARISON WITH SIMULATIONS}
\label{sec:simulations}
\subsection{Cosmological and Galaxy-Scale Simulations}
\label{sec:simulations:cosmological}
Theoretical works have investigated LyC escape using cosmological volumes and on scales of individual star-forming clouds, with most cosmological simulations focusing on LyC in the high-redshift universe. While these studies corroborate some of the observed \fesc\ correlations, their different range of physical conditions and more realistic ISM geometries demonstrate that some aspects of LyC escape may be more complex than observed $z\sim0.3$ correlations. Consistent with the observations, cosmological zoom-in simulations agree that \fesc\ depends sensitively on viewing angle, bursty star formation episodes, and stellar feedback injection \citep[\eg][]{trebitsch17, barrow20, katz20}. In these simulations, however, SN feedback regulates the dispersal of birth clouds and introduces a delay of $\sim3.5-10$ Myr between peak star formation and high \fesc\ \citep[\eg][]{trebitsch17, ma20, katz20, barrow20}. This lag naturally results from the adopted timing of mechanical feedback injection and assumptions about radiative feedback effects and cloud geometry on sub-grid scales (e.g., \citealt{trebitsch17}; Section \ref{sec:simulations:clouds}). Delayed SN feedback at low metallicity \citep[\eg][]{jecmen23} could lengthen this offset further. Some simulations emphasize the role of an extended star formation history, with the cumulative effects of past SNe shaping the galaxy-wide ISM geometry and conditions for \fesc\ \citep[\eg][]{barrow20}. In a variation on the two-stage starburst scenario, high-resolution simulations by \citet{ma20} suggest that newly formed stars on the edges of a SN-driven superbubble could optimize LyC escape. 

Some simulations do find a connection between radiative feedback and LyC escape and support the possibility that different feedback mechanisms may regulate \fesc. In simulated dwarf galaxies, both radiative feedback at early times and later SN-driven blowouts can lead to peaks in \fesc\ \citep{kimm17}. High star formation efficiencies and strong \lya\ radiation pressure could further enhance escape at early times \citep{kimm17, kimm18}. In idealized disk galaxies with GP-like ($\sim 15$\%\ \Zsol) metallicities, radiative feedback permits LyC escape at young ($\sim 2$ Myr) ages \citep{yoo20}. Conversely, in more metal-rich disk galaxies, SN feedback aids LyC escape on small scales but can also increase the \hi\ thickness of the galactic disk. This less advantageous impact of SNe bears some resemblance to the neutral outflows observed in weak $z\sim0.3$ LCEs \citep{carr24}; however, these simulations probe lower \sigsfr\ and may not sufficiently resemble the LzLCS+ sample. 

Simulations support some proposed indirect diagnostics but disagree with others. Cosmological radiation hydrodynamical (RHD) simulations validate the \mgii\ diagnostic in some circumstances, specifically in low-metallicity, dust-poor galaxies \citep{katz22b}. Likewise, \fesclya\ tracks \fesc\ \citep{choustikov24b} and matches the observed low-redshift trend. Other \lya\ properties, like EW and $v_{\rm sep}$, appear necessary but insufficient indicators of \fesc\ \citep{choustikov24b, giovinazzo24, yuan24}, but the simulated relationships do not fully overlap the observations. 

RHD zoom-in simulations of a single galaxy observed from different times and viewing angles demonstrate how observed ISM lines arise from the line-of-sight geometry. For this simulated galaxy, the \cii\ residual flux sets a lower limit on \fesc, but the translation between \cii\ and \fesc\ is not as simple as the picket fence model \citep{mauerhofer21}. The ratio of \cii\ to \hi\ column densities varies spatially \citep{mauerhofer21}, and the observed continuum emission and residual flux in LIS lines can even originate from separate spatial locations \citep{gazagnes23}. This added complexity may explain some of the observed scatter in these diagnostics. Contrary to observations, the simulations do not support \hi\ absorption lines as a diagnostic, but they also do not reproduce the full range of observed Lyman-$\beta$ profiles \citep{mauerhofer21}. Similarly, the simulated \lya\ profiles have complex origins, with features set by diverse gas components and $v_{\rm sep}$ strongly affected by the extended circumgalactic medium (CGM). However, certain GP \lya\ profiles have no match in the simulations, which may suggest that the gas geometry of strong LCEs is not adequately captured by this particular galaxy \citep{blaizot23}. 

Cosmological simulations imply that the association between O32 and \fesc\ is coincidental, without the tight correlation seen in observations. The \fesc\ peaks shortly after the peak in O32, when O32$\sim$3-10, as LyC escape occurs after SNe disrupt clouds  \citep{barrow20, katz20, choustikov24a}. Using a cosmological RHD simulation, \citet{choustikov24a} develop a multivariate model for predicting \fesc. Like the low-redshift observations, they find that \fesc\ scales with $\beta$ and, albeit weakly, with \sigsfr. Because the galaxies only exhibit high \fesc\ post-SNe, their model incorporates an anti-correlation with O32. This model disagrees with predictions from empirical low-redshift diagnostics (Figure \ref{fig:highz}) and does not work well at predicting \fesc\ for the LzLCS+ sample \citep{jaskot24b}; however, the LzLCS+ based diagnostics would likely fail to predict \fesc\ for these simulated galaxies, since these samples show distinct properties.

The disagreements between the simulated and observed trends with \fesc\ may stem from a difference in the sampled populations, missing physical details in the simulations, systematic differences between the real and synthetic observations, or a combination of these factors. The simulated galaxies genuinely differ from strong, low-redshift LCEs, with simulated systems representing galaxies undergoing less extreme starbursts than the $z\sim0.3$ samples, especially the GPs \citep{mauerhofer21, katz22b, choustikov24a}. Although nearby analogs seem similar to observed $z>6$ galaxies (Section \ref{sec:highz}), both the low-redshift starbursts and these first JWST samples may represent the more extreme end of the galaxy population. The  simulations and observations also conflict on the relationship between age and \fesc, with the simulations failing to reproduce \fesc\ for the high O32 and young burst ages associated with the strongest low-redshift LCEs. This particular discrepancy could be a resolution effect, as higher-resolution cloud-scale simulations demonstrate that clusters can emerge from their birth clouds prior to SNe. 

\subsection{Simulations of Star-Forming Clouds}
\label{sec:simulations:clouds}
In contrast to cosmological simulations, higher-resolution (0.02-0.25 pc), cloud-scale simulations identify several factors that can cause substantial LyC escape at young ages. Specifically, \fesc\ depends on the gas morphology and the competition between the radiation field and the gas surface density. Turbulent and inhomogeneous gas structures within molecular clouds play an essential role in LyC transmission \citep[\eg][]{kimm19, kakiichi21}, and the 10 pc resolution typical of many cosmological simulations may thus miss LyC escape at early times ($<6$ Myr; e.g., \citealt{kimm19}). Turbulence creates a high density contrast between dense clumps or filaments and low \nhi\ channels, and \fesc\ can fluctuate dramatically due to the ever-changing gas configuration \citep{kakiichi21}. The \lya\ profiles emerging from the channels do indeed appear narrow at times of high \fesc\ \citep{kimm19, kakiichi21} but need additional scatterings in the wider ISM and CGM to match observations \citep{kimm19}. 

At young ages, radiative feedback dominates the cloud evolution, clearing the lower \nhi\ channels created by turbulence \citep{kakiichi21, menon24}. Radiation can disperse gas well before SNe explode ($<3$ Myr) but dense pockets of gas remain, consistent with detections of such material in GPs (e.g., \citealt{dale12, kimm22, menon24}). Lower metallicities, a higher upper limit for the stellar initial mass function, or higher star formation efficiencies can enhance the effect of radiative feedback and strongly affect the timing and strength of the emergent LyC \citep[\eg][]{kimm19, kimm22, menon24}. Radiation pressure from \lya\ can also increase \fesc\ by an order of magnitude and allow photons to escape at earlier ages (2 Myr), even when stars are co-located with dense gas \citep{kimm19}. Many of these factors could be at play in the super star clusters hosted by strong LCEs, and simulations of star formation and feedback under such extreme conditions could give valuable insight into the physics of LyC escape.

While these simulations demonstrate that LyC can escape from star clusters at young ages, consistent with the observed low-redshift LCEs, neutral gas at larger scales will also affect photon propagation \citep[\eg][]{yoo20}. In a possible conflict with observations, the cloud-scale simulations find a consistent increase of \fesc\ with time \citep[\eg][]{kimm19, menon24}, whereas the observations seem to suggest lower \fesc\ for slightly older populations (Section \ref{sec:feedback:mechanical}). Later SN-driven outflows could interact with dense gas outside the cloud simulation volume \citep[\eg][]{yoo20}, neutral gas could reform outside the volume, or the older bursts in the low-redshift observational samples differ in some unknown way from the younger ones. If \fesc\ does indeed remain high at all times, it may exacerbate the reionization LyC budget problem \citep{munoz24}. 

Cosmological and cloud-scale simulations highlight different properties that control \fesc. The global ISM geometry, shaped by generations of SNe, and the extended CGM gas can affect LyC propagation and observable diagnostics. At the same time, sub-parsec structures may be key to LyC escape from star-forming regions, especially at the ages of peak LyC production. For a comprehensive picture, we need to link small scale processes in molecular clouds with galactic-scale processes such as mergers and galactic winds. 

\section{OPEN QUESTIONS}
\label{sec:questions}
Despite dramatic progress in characterizing the properties of low-redshift LCEs, many gaps remain in our understanding of the nature of cosmic reionizers. To properly connect low-redshift insights to high-redshift observations, we need to understand which LCE properties are universal and which evolve. Are low-redshift LCEs truly analogs of $z>6$ galaxies? Where do these comparisons break down and what impact does that have on inferring \fesc\ at $z>6$? The existence of high-redshift galaxies without low-redshift analogs illustrates that physical conditions at high and low redshift can differ dramatically. In order to test low-redshift \fesc\ diagnostics, we need a census of these same properties in $z\sim3$ LCEs, which will be provided by rest-frame optical JWST observations. Future JWST observations will also reveal the ways in which the $z>6$ population resembles or differs from putative low-redshift analogs. Informed by these data, LyC observations should push beyond the parameter space set by current $z\sim0.3$ surveys (Figure \ref{fig:history}), exploring \fesc\ in fainter, lower sSFR, brighter, more compact, or otherwise more diverse systems. Future missions with FUV capabilities (e.g., the NASA Habitable Worlds Observatory; HWO) may constrain LyC escape in new low-redshift candidates and expand on current efforts. Indirect \fesc\ estimates can also be validated against complementary observations of the epoch of reionization, as any proposed reionizing population should match constraints on the evolving IGM optical depth and the inferred sizes of ionized bubbles around early galaxies.

Both observations and simulations suffer from a disconnect between the physics of LyC escape on large versus small scales. Most low-redshift LyC detections are at $z\sim0.3$, where spatial information is coarse. The ISM structure emerges in greater detail for nearby galaxies, yet their LyC estimates are uncertain or inferred indirectly. Likewise, simulations of entire galaxies may miss crucial fine-scale inhomogeneities that regulate \fesc\ in star-forming regions, while simulations of the latter lack information on the propagation of LyC photons in the broader ISM (Section \ref{sec:simulations}). We need a better understanding of how global galaxy properties, such as those measured in $z\sim0.3$ galaxies and in $z>6$ JWST surveys, are connected to the local sites of LyC production and escape. Do such sites dominate certain global emission properties? Do they reside in particularly favorable locations, shaped by past galaxy interactions or feedback? Why do simulations and $z\sim0.3$ observations disagree regarding the ages and properties of strong LCEs? How do super star clusters evolve and interact with the surrounding ISM at early ages, prior to SN onset? Future spatially-resolved observations of LCEs at LyC, \lya, and other wavelengths can reveal how LyC escape is connected to galaxy structure and local geometry. Follow-up imaging of LzLCS+ galaxies on scales of hundreds of parsecs is underway (the HST Lyman-alpha and Continuum Origins Survey, PI: Hayes). Future UV missions that can reach bluer wavelengths than HST/COS could enable LyC observations for even closer galaxies, whose structure and properties could be studied on smaller scales. 

Current observations offer tantalizing hints as to processes that may be relevant to LyC escape but which require additional investigation. Do different feedback mechanisms or galaxy properties regulate LyC escape in different populations, and if so, where or when do they operate? Do radiative and mechanical feedback operate in tandem, as suggested by some resolved observations? Are galaxy mergers omnipresent in LCEs, and what is their role at high redshift? Due to their distance, most known LCEs lack \hi\ 21 cm observations, which limits our knowledge of the geometry and amount of absorbing gas. X-ray observations of LCEs have also been sparse. We need to establish the prevalence of X-ray sources in LCEs and determine whether and how X-ray sources are causally connected to LyC-emitting sites. The far-ultraviolet LyC regime also remains under-explored but is essential to determine how a galaxy's 900 \AA\ output translates to its total emergent LyC. 

The physics of LyC escape is closely tied to many fundamental yet uncertain aspects of galaxy evolution. Star formation efficiency, both galaxy-wide and at star cluster scales, will determine the progression of star formation as well as the balance between LyC photons and their enshrouding gas. By setting a galaxy's quota of its most massive stars, the IMF will significantly impact its LyC production and sources of feedback. At the low metallicities common in $z\sim0.3$ LCEs and at $z>6$, many relevant properties of massive stars also remain unknown, from the shape of their ionizing spectra to the timing and strength of the mechanical feedback that may aid LyC escape \citep[\eg][]{eldridge22}. Future advances in any of these areas will inform our understanding of both galaxy evolution and the ionization of the IGM.

\begin{summary}[SUMMARY POINTS]
\begin{enumerate}
\item LyC observations at $z\lesssim0.5$ are immune from IGM interference and can provide key information about the physical processes that control LyC escape and the variation of \fesc\ across galaxy populations.
\item The observed \fesc\ in $z\sim0.3$ galaxies correlates strongly with tracers of the line-of-sight ISM absorption from neutral gas and dust. High \fesc\ is also closely associated with high nebular ionization and concentrated star formation, properties which reflect the role of feedback in LyC escape.
\item Multi-wavelength observations of nearby LyC-leaking candidates reveal ISM geometry and feedback effects in detail. Multi-stage starbursts, efficient star formation, X-ray sources, and galaxy interactions may all contribute to LyC escape. Indirect signatures of LyC escape may not originate from the same spatial regions.
\item The strongest LCEs may have escape in most directions, although the exact \fesc\ varies along different sight lines. 
\item Strong LCEs contain young ($< 3$ Myr old) populations, and radiative feedback may dominate the clearing of neutral gas. In weaker LCEs, SN-driven outflows from older populations may clear channels for anisotropic LyC escape.
\item Indirect diagnostics provide a tool for predicting \fesc\ in the epoch of reionization but caution is warranted. Correlations with single variables have high scatter, different multivariate methods can disagree, and predictions sometimes involve extrapolating outside the observed low-redshift parameter space. 
\item Simulations show that the relationship between \fesc\ and observable diagnostics can be complex. Galaxy-scale simulations typically find that SN feedback and slightly older ages are required for high \fesc. In contrast, higher-resolution simulations of star-forming clouds show that small-scale inhomogeneities, radiative feedback, and high star formation efficiency can enable high \fesc\ at younger ages, coincident with high LyC production. These latter processes may drive the high \fesc\ observed in young, compact starbursts at low redshift.
\end{enumerate}
\end{summary}

\begin{issues}[FUTURE ISSUES]
\begin{enumerate}
\item The redshift evolution of indirect \fesc\ diagnostics remains uncertain. We need to determine the ways high-redshift galaxies may differ from low-redshift LCEs.
\item Despite considerable progress in LyC detections, the full parameter space has yet to be investigated. LyC constraints in faint galaxies are especially important for understanding reionization.
\item We need a theoretical and observational understanding of the physics of LyC escape on small scales and how it relates to global galaxy properties and global \fesc.
\item Certain multi-wavelength properties of LCEs remain under-explored, including the total \hi\ content, contribution of X-ray sources, and spectral shape of the LyC.
\item Research on the evolution, mass function, and feedback contribution of low-metallicity massive stars will aid our understanding of LyC production and escape. 
\end{enumerate}
\end{issues}

\section*{DISCLOSURE STATEMENT}
The authors are not aware of any affiliations, memberships, funding, or financial holdings that
might be perceived as affecting the objectivity of this review. 

\section*{ACKNOWLEDGMENTS}
AJ thanks Sophia Flury, G{\"o}ran {\"O}stlin, and Maxime Trebitsch for reading and providing insightful comments on portions of this article. Thank you to Sara Mascia and Xinfeng Xu for sharing data and to John Chisholm, Sophia Flury, Ryan Keenan, Lena Komarova, Alexandra Le Reste, {\'A}ngel L{\'o}pez-S{\'a}nchez, Genoveva Micheva, Sally Oey, G{\"o}ran {\"O}stlin, Andrea Prestwich, and Alberto Saldana-Lopez for permission to use figures. Thank you to the entire LzLCS team for thoughtful discussions over the past few years; these exchanges have shaped many of the conclusions in this article. I dedicate this review to the memory of Professor Karen Kwitter, who sparked my interest in spectroscopy, the ISM, and teaching.

\noindent

\end{document}